\documentclass[aip,pop,amsmath,amssymb,reprint,balance,superscriptaddress]{revtex4-1}
\usepackage{graphicx}
\usepackage{dcolumn}
\usepackage{bm}
\usepackage{color}
\usepackage{soul}
\usepackage{hyperref}
\hypersetup{
    colorlinks,%
    citecolor=blue,%
    filecolor=blue,%
    linkcolor=blue,%
    urlcolor=blue
} 

\begin{document}

\title {Anisotropic energy transfers in  quasi-static magnetohydrodynamic turbulence}
\vskip 0.3 in

\author{K. Sandeep Reddy}
\email{ksreddy@iitk.ac.in}
\affiliation{Department of Mechanical Engineering, Indian Institute of Technology, Kanpur 208016, India}

\author{Raghwendra Kumar }
\email{raghav@barc.gov.in}
\affiliation{Theoretical Physics Division, Bhabha Atomic Research Centre, Mumbai 400 085, India}

\author{Mahendra K. Verma}
\email{mkv@iitk.ac.in}
\affiliation{Department of Physics, Indian Institute of Technology, Kanpur 208016, India}

\date{\today}
\vskip 0.2 in

\begin{abstract}
We perform direct numerical simulations of quasi-static magnetohydrodynamic turbulence, and compute various energy transfers including the ring-to-ring  and conical energy transfers, and the energy fluxes of the perpendicular and parallel components of the velocity field.  We  show that the rings with higher polar angles transfer energy to ones with lower polar angles.  For large interaction parameters, the dominant energy transfer takes place near the equator (polar angle $\theta \approx \frac{\pi}{2}$).  The energy transfers are local both in wavenumbers and angles.  The energy flux of the perpendicular component is predominantly from  higher to lower wavenumbers (inverse cascade of energy), while that of the parallel component is from lower to higher wavenumbers (forward cascade of energy). Our results are consistent with  earlier results, which indicate  quasi  two-dimensionalization of quasi-static magnetohydrodynamic (MHD) flows at high interaction parameters.
\end{abstract}
\maketitle


\section{Introduction}
\label{sec:intro}
Liquid-metal flows under strong magnetic field occur in  geophysics, metallurgical applications like metal-plate rolling,  heat exchangers of the proposed fusion reactor ITER, etc.  These flows are described by magnetohydrodynamics (MHD), which involves equations for the velocity and magnetic fields.   Liquid metals  have small magnetic Prandtl numbers $\mathrm{Pm}$, which is the ratio of the kinematic viscosity $\nu$ to the  magnetic diffusivity $\eta$.\cite{Roberts:book,Knaepen:ARFM2008} 

The flow velocity in a typical industrial application is rather small.  Hence  the magnetic Reynolds number $\mathrm{Rm}$ ($UL/\eta$, where  $U$ and $L$ are the large-scale velocity and length scales respectively)  for such flows is quite small.  A limiting case of such flows,  called the quasi-static limit\cite{Roberts:book,Knaepen:ARFM2008} ($\mathrm{Rm} \rightarrow 0$), provides further simplification; here the time derivative of the magnetic field is negligible compared to the magnetic diffusion term.  
Experiments\cite{Alemany:JMec1979,Kolesnikov:FD1974} and numerical simulations\cite{Schumann:JFM1976,Zikanov:JFM1998,Favier:POF2010b} show that the flow becomes quasi-two-dimensional when subjected to a strong mean magnetic field. In the present paper, we discuss the energy transfers in the quasi-static MHD.  We highlight the energy transfers responsible for making the flow quasi two-dimensional.

The external magnetic field makes the flow anisotropic. For a strong magnetic field, Moffatt\cite{Moffatt:JFM1967} predicted a rapid decay of isotropic three-dimensional turbulence to a two-dimensional state. Kit and Tsinober\cite{Kit:MG1971} analyzed several experimental results and argued that MHD flow under strong magnetic field is two-dimensional.  Alemany {\it et al.}~\cite{Alemany:JMec1979} performed experiment on mercury and obtained a $k^{-3}$ energy spectrum.  Alemany {\it et al.}~\cite{Alemany:JMec1979} and Moreau,\cite{Moreau:book} however, explained this spectrum by arguing that the nonlinear transfer time is independent of the wavenumber $k$, not due to the two-dimensionality of the flow; they proposed that the quasi-static MHD is quasi two-dimensional.  Sommeria and Moreau\cite{Sommeria:JFM1982} studied conditions when the MHD turbulence at low-$\mathrm{Rm}$ becomes two-dimensional. Klein and Poth\'{e}rat,\cite{Klein:PRL2010} and Poth\'{e}rat and Klein\cite{Potherat:ARXIV2014} studied the three dimensionalization of wall-bounded MHD flows in a quasi two-dimensional flow of liquid metals; these works as well as Poth\'{e}rat\cite{Potherat:MHD2012} emphasize the role of boundary walls in the dynamics of quasi-static MHD.

The aforementioned quasi two-dimensionalization has been studied using direct numerical simulations.  Burattini {\it et al.}\cite{Burattini:PD2008,Burattini:PF2008}  computed  the kinetic energy spectrum and  showed how the anisotropy varies with respect to the direction of the external magnetic field.  Favier {\it et al.}~\cite{Favier:POF2010b,Favier:JFM2011} studied this phenomena using  direct numerical simulations (DNS) and eddy-damped quasi-normal Markovian (EDQNM) model.  Zikanov and Thess\cite{Zikanov:JFM1998} showed that for moderate interaction parameters, the turbulence remains quasi two-dimensional for several eddy turnover times  before  it is interrupted by strong bursts of three dimensional turbulence.    Reddy and Verma\cite{Reddy:POF2014} quantified the energy distribution using ring spectrum, and show that the energy is concentrated near the equator.  They also showed that they energy spectrum is exponential ($\exp(-bk)$) for a very large magnetic field. 
 
The above simulations, performed using pseudo-spectral method in a periodic box, capture the properties of the bulk flow quite well.  For example, steepening of the energy spectrum with the increase of interaction parameter is observed in all the simulations\cite{Burattini:PD2008,Reddy:POF2014} as well as in experiments.\cite{Branover:PTR1994,Eckert:HFF2001}  However, the Hartmann layers  cannot be studied using periodic box simulations.   Dymkou and Potherat,\cite{Dymkou:TCFD2009} and Kornet and Potherat\cite{Kornet:ARXIV2014} have developed numerical techniques to simulate wall bounded MHD flows using least dissipative modes. 
Boeck {\it et al.}\cite{Boeck:PRL2008} performed DNS of quasi-static MHD flow in a channel with no-slip walls and observed recurring transitions between two-dimensional  and three-dimensional states in the flow. 

However, a word of caution is in order.  Most of the aforementioned simulations have been performed on a periodic box.  The flow structures with realistic boundary conditions (e.g. no-slip walls) differ significantly from those with periodic domains, since boundary effects  are completely ignored in periodic box simulations.   In a wall-bounded low-Rm liquid-metal MHD flow, the Hartmann layers at walls restrict the elongation of two-dimensional structures; these features are not captured in periodic box simulations.  The structures longer than the length of the domain are cut at the periodic boundaries and appear as 2D structures.

Yet, periodic box computations provide interesting insights into energy transfers in the bulk flow.  The energy spectrum computed using the periodic box simulations are in general agreement with those computed in experiments, for example, quasi two-dimensionalization of the flow is captured  successfully in periodic box simulations.~\cite{Zikanov:JFM1998} 

The energy spectrum of liquid-metal flows has been studied by a large number of scientists and engineers (see above).  However, diagnostics like energy flux, shell-to-shell energy transfer, etc. are much less  studied in this field.  In fluid turbulence, the turbulence is homogeneous and isotropic in the inertial range.~\cite{Lesieur:Book}  Also, in the inertial range, Kolmogorov's flux is constant, and the shell-to-shell energy transfer is forward and local (maximum transfers between the neighboring shells).~\cite{Lesieur:Book}  However, in liquid-metal flows,  the mean magnetic field  induces anisotropic energy transfers, which are quantified using the angular-dependent energy flux and ring-to-ring transfers.   We use the formalism proposed by Dar {\it et al.,}\cite{Dar:PD2001} Verma,\cite{Verma:PR2004} and Teaca {\it et al.}\cite{Teaca:PRE2009} to compute these quantities.   

\begin{figure}[htbp]
\begin{center}
\includegraphics[width=4.4cm,angle=0]{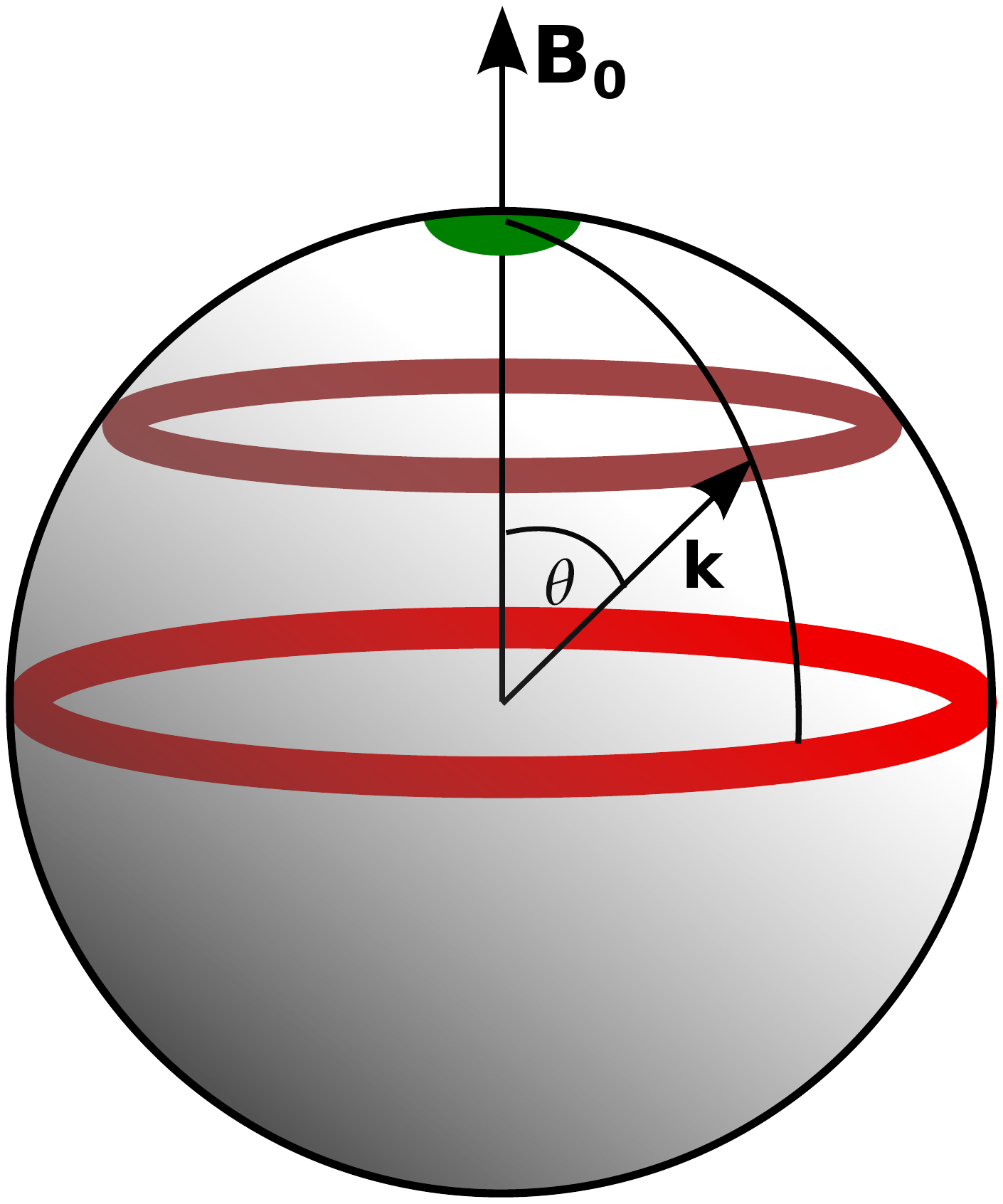}
\end{center}
\caption{Ring decomposition of the Fourier space.\cite{Teaca:PRE2009,Reddy:POF2014}}
\label{fig:ring-decompose}
\end{figure}

\begin{figure}[htbp]
\begin{center}
\includegraphics[scale=0.75]{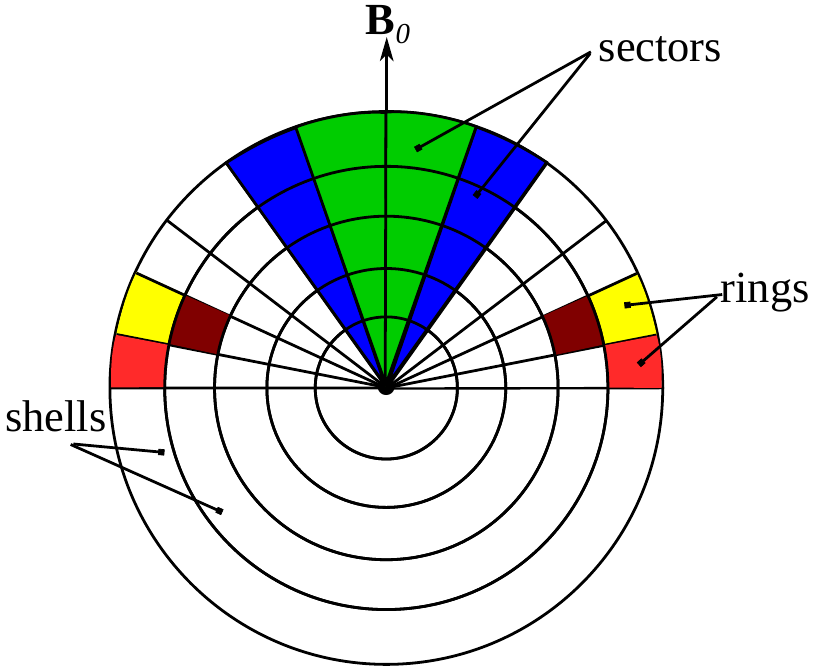}
\end{center}
\caption{A cross-sectional view of wavenumber shells, sectors, and rings.}
\label{fig:shell_sec_ring}
\end{figure}

For magnetohydrodynamic flows with unit magnetic Prandtl number,  Teaca {\it et al.}\cite{Teaca:PRE2009} computed the energy transfers among the spectral rings (see Fig.~\ref{fig:ring-decompose}).  These rings are specified by their radii and  sector indices (see Fig.~\ref{fig:shell_sec_ring}). For convenience, we refer to the rings near the pole  as ``polar rings'' ($\theta \approx 0$), and those near the equator as ``equatorial rings'' ($\theta \approx \pi/2$). In this paper, we compute the energy transfers among the rings, and show that  the energy transfers are dominant near the plane perpendicular to the external magnetic field when the external field is large.  We also compute other quantities, like, the energy flux, conical energy flux, and ring dissipation rates.   These results provide newer insights into the quasi-two-dimensional nature of quasi-static MHD turbulence at  high interaction parameters.\cite{Zikanov:JFM1998,Favier:POF2010b,Reddy:POF2014}  Note, however, that our work differs from that of Favier {\it et al.}~\cite{Favier:POF2010b,Favier:JFM2011}.  We explicitly compute the energy transfers (in contrast to Favier {\it et al.}~\cite{Favier:POF2010b,Favier:JFM2011} who focus on the energy spectra of the poloidal and toroidal components), anisotropy of the flow, as well the nonlinear transfer spectrum.

The paper is organized as follows: In Sec.~\ref{sec:theory}, we present the formalism of ring-to-ring energy transfers, conical energy flux, and parallel and perpendicular energy fluxes. Section~\ref{sec:sim} contains the details of our numerical simulations. We present the results of our numerical computations in Sec.~\ref{sec:results}, and summarize the results in Sec.~\ref{sec:summary}.

\section{Theoretical Framework}
\label{sec:theory}

\subsection{Governing equations}
The governing equations of low-Rm liquid-metal flows under quasi-static approximation are:\cite{Roberts:book,Knaepen:ARFM2008}
\begin{eqnarray}
\frac{\partial{\bf u}}{\partial t} + ({\bf u}\cdot\nabla){\bf u} &=& -\nabla{(p/\rho)} - \frac{\sigma {B_0}^2}{\rho} \Delta^{-1} \frac{\partial^2{\bf u}}{\partial z^2}\nonumber \\
 && + \nu\nabla^2 {\bf u} + {\bf f}, \label{eq:NS}\\
\nabla \cdot {\bf u} & = & 0, \label{eq:continuity} 
\end{eqnarray}
where {\bf u} is the velocity field, ${\bf B_0}=B_0 \hat{z}$ is the  constant external magnetic field, $p$ is the pressure, $\rho$, $\nu$, $\sigma$ are the density, kinematic viscosity, conductivity of the fluid, respectively, $\Delta^{-1}$ is the inverse of the Laplacian operator, and ${\bf f}$ is the forcing.   We also assume that the flow is incompressible, i.e., the density of the fluid is constant.

The above equations are nondimensionalized using  the characteristic velocity $U_0$ as the velocity scale, the box dimension $L_0$ as the length scale, and $L_0/U_0$ as the time scale.  As a result, the non-dimensional equations are
\begin{eqnarray}
\dfrac{\partial{\bf U}}{\partial T} + ({\bf U}\cdot\nabla'){\bf U} &=& -\nabla'{P} - B^{\prime 2}_0 \Delta^{\prime -1} \dfrac{\partial^2{\bf U}}{\partial Z^2} \nonumber \\
&& + \nu^\prime \nabla^{\prime 2} {\bf U} + {\bf f^\prime}, \label{eq:NS2}\\
\nabla^\prime \cdot {\bf U} & = & 0, \label{eq:continuity2} 
\end{eqnarray}
\noindent
where non-dimensional variables are $\mathbf U = \mathbf u/U_0$, $ \nabla' = L_0 \nabla $, $\Delta^{\prime -1} = \Delta^{-1}/L_0^2$, $T = t(U_0/L_0)$, 
$B_0^{\prime 2}  = \sigma B_0^2 L_0 /(\rho U_0)$, and $\nu^\prime=\nu/(U_0 L_0)$.  

In quasi-static MHD turbulence, there is an interplay between the Joule dissipation, viscous dissipation, and the non-linear energy transfers at various scales.   It is convenient to analyze the  aforementioned processes in the wavenumber or the Fourier space. The non-dimensional  equations in the spectral space\cite{Schumann:JFM1976,Zikanov:JFM1998,Knaepen:JFM2004} are
\begin{eqnarray}
\dfrac{\partial{\hat{U}_i(\bf{k})}}{\partial T} &=& -ik_j \sum \hat{U}_j({\bf q}) \hat{U}_i({\bf{k}-\bf{q}}) - ik_i \hat{P}({\bf k}) \nonumber \\
&-& {B'_0}^2{\mathrm{cos^2}}(\theta)\hat{U}_i({\bf k}) - \nu' k^2 \hat{U}_i({\bf k})+\hat{f_i^{\prime}}({\bf k}), \label{eq:NS-fourier} \\
k_i  \hat{U}_i(\mathbf{k}) & = & 0,
\end{eqnarray}
where $\hat{U}_i(\mathbf{k})$,  $\hat{P}(\mathbf{k})$,  and $\hat{f_i^{\prime}}({\bf k})$ are the Fourier transforms of the velocity, pressure,  and force fields, respectively, and $\theta$ is the angle between wavenumber vector ${\bf k}$ and the external magnetic field $\mathbf B_0$.

The Reynolds number, which is the ratio of the nonlinear term to the viscous term, is  a measure of nonlinearity in the flow.  The  interaction parameter, which is the ratio of the Lorentz force to the nonlinear term, quantifies the strength of the Lorentz force.  The interaction parameter $N$ is defined as 
\begin{equation}
N= \frac{{B_0'^{2}} L}{U^{'}},
\label{eq:int_param}
\end{equation}
\noindent
where $U^{'}$ is the root mean square (rms) of the velocity  defined\cite{Vorobev:POF2005,Burattini:PD2008} as
\begin{equation}
\frac{3}{2}U'^2=E = \int_0^{\infty}E(k)dk,
\end{equation}
and  $L$ is the the non-dimensional integral length scale defined as
\begin{equation}
L= {\frac{\pi}{2 U^{'2}}}{\int_0^{k_{max}}{\dfrac{E(k)}{k}dk}},
\label{eq:int_length}
\end{equation}
where $E(k)$ is the one-dimensional energy spectrum.  The energy equation corresponding to Eq.~(\ref{eq:NS-fourier}) is
\begin{eqnarray}
{{\partial E({\bf k})}\over {\partial T}} &=&  T({\bf k})-2 {{B'}_0^{2}}{\cos^{2}(\theta)}E({\bf k})\nonumber \\
&&- 2 \nu' k^2 E({\bf k})+F({\bf k}), \label{eq:energy}
\end{eqnarray}
\noindent 
where $E({\bf k})=|{\bf \hat{U}}({\bf k})|^2/2$, $F({\bf k})$ is energy supply rate due to external forcing $\mathbf f'$, and $T({\bf k})$ is the net nonlinear energy transfer rate to a mode ${\mathbf k}$.   The energy equation contains two dissipative terms: the Joule dissipation rate
\begin{equation}
\epsilon_J({\bf k})  = 2 {{B^\prime_0}^{2}}\cos^{2}(\theta)E({\bf k}),\label{eq:JD}
\end{equation}
and viscous dissipation rate
\begin{equation}
\epsilon_\nu({\bf k})  = 2 \nu' k^2 E({\bf k}).  \label{eq:VD}
\end{equation}

The nonlinear interactions among the Fourier modes yield energy transfers among the modes. We quantify these transfers using energy flux, shell-to-shell and  ring-to-ring energy transfers, etc. which will be described below.

\subsection{Shell-to-shell and  ring-to-ring energy transfers, and conical energy flux}\label{subsec:energy_tr}
We can study the  energy transfers in the Fourier space in detail using the ``mode-to-mode'' energy transfer proposed by Dar {\it et al.}\cite{Dar:PD2001} and Verma.\cite{Verma:PR2004}  For a triad ${(\bf k, \bf p, \bf q)}$, 
\begin{equation}
S({\bf k|p|q}) = \mathrm{\Im} \{ [ {\bf k \cdot {\hat U}(q)] [\hat U^*(k) \cdot \hat U(p)}] \},\label{eq:mode-mode}
\end{equation}
is the mode-to-mode energy transfer rate from the mode ${\mathbf p}$ to the mode ${\mathbf k}$ with the mode ${\mathbf q}$ acting as a mediator.\cite{Dar:PD2001,Verma:PR2004} Here, $\Im$ and * represent the imaginary part and the complex conjugate of a complex number, respectively.  Note that $ {\bf k} = {\bf p}+{\bf q}$.
 
The shell-to-shell energy transfer rate from all the modes in the  $m^{\mathrm{th}}$ shell to the modes in the $n^{\mathrm{th}}$ shell is defined as 
\begin{equation}
T^m_n=\sum_{{\bf k} \in n}\sum_{{\bf p} \in m}S({\bf k|p|q}).
\end{equation}
The shell-to-shell energy transfer provides an average energy transfer  over all angles.  To diagnose the angular dependence of the energy transfer, we divide the wavenumber shells into rings, as shown in Fig.~\ref{fig:ring-decompose}. A ring is an intersection of a shell and a sector (see Fig.~\ref{fig:shell_sec_ring}), hence it is characterized by $(m,\alpha)$, where $m$ denotes the shell index, and $\alpha$ represents the sector index.       The ring-to-ring energy transfer rate from the ring $(m,\alpha)$ to the ring $(n,\beta)$ is~\cite{Teaca:PRE2009}
\begin{equation}
T_{(n,\beta)}^{(m,\alpha)}=\sum_{{\bf k} \in (n,\beta)} \sum_{{\bf p} \in (m,\alpha)}  S({\bf k}|{\bf p}|{\bf q}).
\end{equation}
The ring-to-ring energy transfers are normalized using $A_{i} = |\mathrm{cos}(\theta_i)-\mathrm{cos}(\theta_{i+1})|$ to compensate for the uneven distribution of modes in the rings.\cite{Teaca:PRE2009}  The rings closer to the equator have more Fourier modes than those near the poles. Hence, we define a  normalized ring energy transfer function as
\begin{equation}
\overline{T}_{(n,\beta)}^{(m,\alpha)} = \frac{1}{A_{\alpha}A_{\beta}}T^{(m,\alpha)}_{(n,\beta)}.
\end{equation}
The properties of the ring-to-ring transfers are listed by Teaca {\it et al}.\cite{Teaca:PRE2009}  For example, the energy transfer rates between rings within a shell vanish for isotropic flows.   In this paper, we  will adopt Teaca {\it et al.}'s\cite{Teaca:PRE2009} procedure for these computations.

\begin{figure}[htbp]
\begin{center}
\includegraphics[width=4.2cm,angle=0]{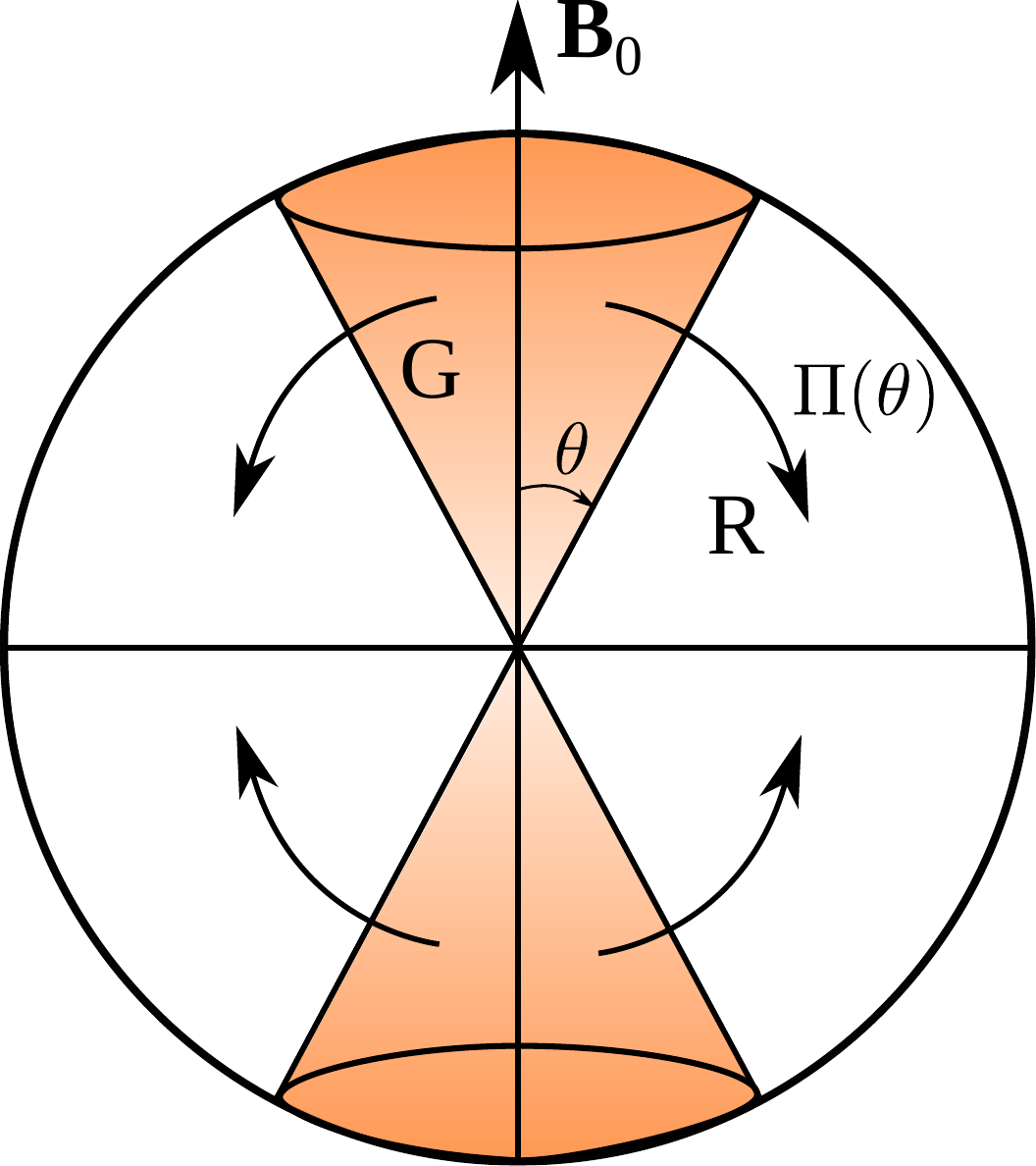}
\end{center}
\caption{ Conical energy flux $\Pi(\theta)$ is the rate of energy transfer from the modes inside a cone of semi-vertical angle $\theta$ to the modes outside the cone (see Sec.~\ref{subsec:energy_tr}). }
\label{fig:flux-cone}
\end{figure}

To further quantify the anisotropic energy transfers, we compute another quantity, called the conical energy flux $\Pi(\theta)$.   Consider a cone of semi-vertical angle $\theta$, as shown in Fig.~\ref{fig:flux-cone}.  The conical energy flux $\Pi(\theta)$ is  defined as the total energy transfer from the modes inside the  cone  to the modes outside the cone (from region $G$ to region $R$ of Fig.~\ref{fig:flux-cone}): 
\begin{equation}
\Pi(\theta)=  \sum_{{\bf k} \in R} \sum_{{\bf p} \in G}  {S({\bf k|p|q)}}.\label{eq:conical-flux}
\end{equation}
 
We also calculate the energy $E_\alpha$ in the sector $\alpha$ as  
\begin{equation}
E_\alpha = \sum_{{\bf k} \in \alpha} \frac{1}{2} |{\bf \hat{U}({\bf k})}|^2.
\end{equation}
For a strong external field, the energy is concentrated near the equatorial regions (perpendicular to the mean magnetic field). 
The energies in the equatorial and non-equatorial sectors  are given by   
\begin{eqnarray}
E_{\mathrm{eq}} = \sum_{ {\bf k} \in  \alpha_{\rm eq}} \frac{1}{2} |{\bf \hat{U}({\bf k}})|^2, \\
E_{\operatorname{non-eq}} = \sum_{ {\bf k} \notin  \alpha_{\rm eq} } \frac{1}{2} |{\bf \hat{U}({\bf k}})|^2, 
\end{eqnarray}
respectively.  Here, $\alpha_{\rm eq}$ represents the equatorial sector, spanning angles  in then range $\left[\frac{7\pi}{15},\frac{\pi}{2}\right]$.

\subsection{Energy exchange between perpendicular and parallel velocity components}   
Another interesting feature of anisotropic flows is  the energy exchange between the perpendicular and parallel components of the velocity field ($U_{\parallel}={\mathbf U} \cdot \hat{z}$ and ${\mathbf U}_\perp ={\mathbf U} - U_{\parallel} \hat{z} $ respectively).   Here, we compute these transfers using the energy fluxes of the parallel and perpendicular components of the velocity field  (see Appendix A).   In brief, the energy equations for the perpendicular and parallel components of the velocity field are
\begin{eqnarray}
{{\partial E_{\perp}({\bf k})}\over {\partial t}} &=&  {\sum_{\mathbf k = \mathbf p+\mathbf q} } S_{\perp}({\bf k}|{\bf p}|{\bf q}) - 2 {{B{'}}_0^{2}}\cos^{2}(\theta)E_{\perp}({\bf k}) + P_{\perp}({\bf k}) \nonumber \\
&&- 2 \nu' k^2 E_{\perp}({\bf k})+ \Re \{{ \bf \hat{f}'_{\perp}(k) \cdot \hat{U}_{\perp}^*(k)} \},\\
{{\partial E_{\parallel}({\bf k})}\over {\partial t}} &=&  {\sum_{\mathbf k = \mathbf p+\mathbf q} } S_{\parallel}({\bf k}|{\bf p}|{\bf q}) - 2 {{B{'}}_0^{2}}\cos^{2}(\theta)E_{\parallel}({\bf k}) + P_{\parallel}({\bf k})\nonumber \\
 & & -2 \nu' k^2 E_{\parallel}({\bf k})+ \Re \{{ \hat{f}'_{\parallel}(k) \hat{U}_{\parallel}^*(k)}\},
\end{eqnarray}
\noindent 
respectively, where $E_{\perp}({\bf k}) =\frac{1}{2}|\hat {\bf U }_{\perp}({\bf k})|^2$ and $E_{\parallel}({\bf k})=\frac{1}{2}|\hat {U}_{\parallel}({\bf k})|^2$ are the energies of the perpendicular and parallel components of the velocity field, respectively, and
\begin{eqnarray}
S_{\perp}({\bf k|p|q})&=& \mathrm{\Im} \{ [ {\bf k \cdot {\hat U}(q)] [\hat U_{\perp}^*(k) \cdot \hat U_{\perp}(p)}]\},\\
S_{\parallel}({\bf k|p|q})&=& \mathrm{\Im} \{ [ {\bf k \cdot {\hat U}({\bf q})}] [\hat U_{\parallel}^*({\bf k})  \hat U_{\parallel}({\bf p})] \}, \label{eq:Spll}\\
P_{\perp}({\bf k}) &=& \Im \{ [{\bf k_{\perp} \cdot \hat U_{\perp}^*(k)} ] \hat{P}({\bf k})\},\label{eq:P_perp}  \\
P_{\parallel}({\bf k}) &=& \Im \{ [ k_{\parallel}  \hat U_{\parallel}^*({\bf k}) ] \hat{P}({\bf k})\}, \label{eq:P_par}
\end{eqnarray}
and $\Re$, $\Im$, $*$ represent the real and imaginary parts, and the complex conjugate a complex number, respectively.  In the above equations we have replaced ${\bf k'}$ and $\hat{{\mathbf U}}({\mathbf k'})$ in the equations of Appendix~A with $-{\bf k}$ and $\hat{{\mathbf U}}^*({\mathbf k})$ respectively.  Also note that Eqs.~(\ref{eq:P_perp},\ref{eq:P_par}) and the condition ${\bf k \cdot   \hat U}(\mathbf k) = 0$ imply that 
\begin{equation}
P_{\perp}({\bf k}) = - P_{\parallel}({\bf k}).
\end{equation} 
We interpret the above result as following. The energy gained by the perpendicular component ${\bf \hat U_{\perp}^*(k)}$ via pressure is equal and opposite to the energy lost by the parallel component.  The magnitude of the transfer to the   parallel component via  pressure is given by Eq.~(\ref{eq:P_perp}).  Thus pressure facilitates energy transfers between the parallel and perpendicular components of the velocity field.  Note that  there is no direct energy transfer between ${\bf \hat U_{\perp}}$ and ${ \hat U_{\parallel}}$.  

The energy flux $\Pi_\perp(k_0)$ for the perpendicular component of the velocity field for a wavenumber sphere of radius $k_0$  is defined as the net energy transfer  from the modes $\mathbf{U}_\perp(\mathbf p)$  residing inside the sphere to the modes $\mathbf{U}_\perp(\mathbf k)$ outside the sphere,  i.e.,
\begin{equation}
\Pi_{\perp}(k_0) =  \sum_{|{\bf k}| \geq k_0 } \sum_{|{\bf p}| < k_0}  {  S_{\perp}({\bf k|p|q)}}.
\end{equation}
A similar formula for the flux of the parallel velocity component, $\Pi_{||}(k_0)$, is
\begin{equation}
\Pi_{\parallel}(k_0) =  \sum_{|{\bf k}| \geq k_0 } \sum_{|{\bf p}| < k_0}  {  S_{\parallel}({\bf k|p|q)}}.
\end{equation} 
We will compute these quantities using  our simulation data.  

In the following section, we describe the details of simulation method employed for the present study.

\section{Details of Numerical Simulations}
\label{sec:sim}
We use pseudo-spectral code {\it{Tarang}}\cite{Verma:Pramana2013} to solve the non-dimensional  quasi-static MHD equations (Eqs.~(\ref{eq:NS2}) and (\ref{eq:continuity2})) in a cubical box on a $256^3$ grid.  Periodic boundary conditions are  applied in all the three directions. We use the fourth-order Runge-Kutta method for time-stepping,  Courant-Friedrichs-Lewy (CFL) condition for calculating time-step ($\Delta t$), and the $3/2$ rule for dealiasing.\cite{Canuto:book,Boyd:book} We start our simulation for $N=0$ using a model energy spectrum\cite{Pope:book} as the initial condition:
\begin{eqnarray}
E(k) = C\epsilon^{2/3}k^{-5/3}f_L(kL)f_{\eta}(k\eta),
\end{eqnarray}
with the Kolmogorov constant $C=1.5$, and the energy supply rate $\epsilon=1.0$. $f_L, f_{\eta}$ are defined as 
\begin{eqnarray}
f_L(kL) &=& \left(\frac{kL}{[(kL)^2+c_L]^{1/2}}\right)^{5/3+p_0},\\
f_{\eta}(k\eta) &=& \exp(-\beta k \eta),
\end{eqnarray}
where $c_L=1.5$, $p_0=2$ and $\beta = 5.2$. The initial phases of the velocity Fourier modes are randomly generated.

In order to achieve a steady-state, the velocity field is randomly forced using a  scheme similar to that followed by Burattini {\it et al.,}\cite{Burattini:PD2008} Vorobev {\it et al.,}\cite{Vorobev:POF2005} and Carati {\it et al.,}~\cite{Carati:POF1994} which is, 
\begin{eqnarray}
{\bf {\hat{f^{'}}}(k)} = \gamma({\bf k}){\bf {\hat U}(k)},\\
\gamma({\bf k}) = \frac{\epsilon_{in}}{n_{f} [{\bf \hat{U}({\bf k})}. {\bf \hat{U}^*({\bf k})}]},
\end{eqnarray}
where $n_{f}$ is total number of modes inside the forcing wavenumber band.  We choose the energy input rate $\epsilon_{\rm in} = 0.016$, and the forcing band as $1 \leq {\bf |k|} \leq 3 $ for the shell-to-shell, ring-to-ring, and conical flux studies.  However, we choose  the forcing band as $8 \leq {\bf |k|} \leq 9 $ with $\epsilon_{\rm in} = 0.072$, for the computation of  the energy fluxes of the parallel and perpendicular components of the velocity field.

\begin{table}[htbp]
\caption{\label{tab:table1}Details of simulations: the constant external magnetic field $B^\prime_0$, forcing band $k_f$, the interaction parameter $N$ computed at steady state, the interaction parameter $N_0$ computed at the instant when external magnetic field is applied, rms velocity $U^\prime$,  eddy turnover time $\tau$, and time averaged $k_{\rm max}\eta$.  }
\begin{ruledtabular}
\begin{tabular}{ccccccc}
$B_0^\prime$		& $k_f$		& $N$	& $N_0$	& $U^\prime$ & $\tau$ &$k_{\rm max}\eta$ \\ 
\hline
2.29		& [1,3]		& 1.7	& 1.0 	& 0.39 	& 0.32	& 2.4		\\
3.60		& [1,3]		& 5.5	& 2.5 	& 0.35 	& 0.43	& 2.8	\\
5.15		& [1,3]		& 11		& 5.0	& 0.39 	& 0.39	& 2.9	\\
6.26		& [1,3]		& 14		& 7.5	& 0.45 	& 0.37	& 2.9	\\
7.28		& [1,3]		& 18		& 10.0	& 0.51 	& 0.33	& 2.8	\\
10.23	& [1,3]		& 27		& 20.0	& 0.65 	& 0.26	& 2.6	\\
25.1	   	& [1,3]		& 130	& $-$	& 0.86 	& 0.21	& 2.4	\\
32.6    & [1,3]		& 220	& $-$	& 0.87 	& 0.21	& 2.4	\\ 		
19.6		& [8,9]		& 100	& 30		& 0.64 	& 0.26	& 2.1	\\
\end{tabular}
\end{ruledtabular}
\end{table}

We choose a fixed value for the non-dimensional viscosity $\nu'$ as $0.00036$, and vary the non-dimensional external magnetic field $B_0'$  to simulate the $N$'s  ranging from 1.7  to 220. The final state of $N=0$ is used as the initial condition for $N=1.7, 5.5, 11, 14, 18$ and $27$, and all the simulations are carried out till a new statistically steady-state is reached. However, for high interaction parameters, i.e., $N=130$ and $220$ we have used the final state of $N=27$ as the initial condition. The interaction parameter $N$ is calculated using the values of $U'$ and $L$ of the steady state.\cite{Reddy:POF2014}   

  For all our simulations, the grid resolution is chosen such that $k_{\mathrm{max}}\eta > 1.4$, where $k_\mathrm{max}$ is the largest wavenumber of the simulation, and $\eta$ is the Kolmogorov length scale.  Hence, the smallest length scale of the flow is larger than the  grid size.\cite{Jimenez:JFM1993,Favier:POF2010b} Thus our simulations are fully resolved.  We refer to Reddy and Verma\cite{Reddy:POF2014} for the details on the grid independence tests. 

We compute the energy transfer rates using the simulation data and the formulas defined in Sec.~\ref{sec:theory}.  For the shell-to-shell energy transfers, we divide the Fourier space into 19 spherical shells. The radii of the first two shells are 4 and 8, and the last two shell radii are 42.5 and $85 = 128 \times 2/3$, with the factor of 2/3 arising due to de-aliasing. The remaining shells are binned logarithmically that yields the shell radii as: 4.0, 8.0, 8.9, 9.9, 10.9, 12.2, 13.5, 14.9, 16.6, 18.4, 20.5, 22.7, 25.2, 28.0, 31.1, 34.5, 38.3, 42.5 and 85.0.   We choose logarithmic binning for the intermediate shells because the energy transfers are local for these shells.  The radii of first two shells are chosen as 4 and 8 since the number of modes is small in these shells.  

For the ring-to-ring and the conical flux transfers, the aforementioned shells are further divided into rings.  In our simulation, we work with the modes with $k_z \ge 0$ or $0 \le \theta \le \pi/2$ by exploiting the reality condition.   We divide the Fourier space into 15 equi-spaced  sectors for $0 \le \theta \le \pi/2$. The range of angles for the $i^\mathrm{th}$ sector is $[\frac{(i-1)\pi/2}{15}, \frac{i\pi/2}{15})$, with $i=1,2,3, \ldots 15$.  

The results of our simulation data are presented in the following section.

\section{Numerical results}
\label{sec:results}
We compute various energy transfer rates for $N=1.7, 5.5, 11, 14, 18, 27, 130$, and $220$.  A detailed description of each transfer is described in the following subsections.

\subsection{Anisotropic energy spectrum}\label{subsec:anisotropic_spectrum}
\begin{figure}[htbp]
\begin{center}
\includegraphics[width=8.3cm,angle=0]{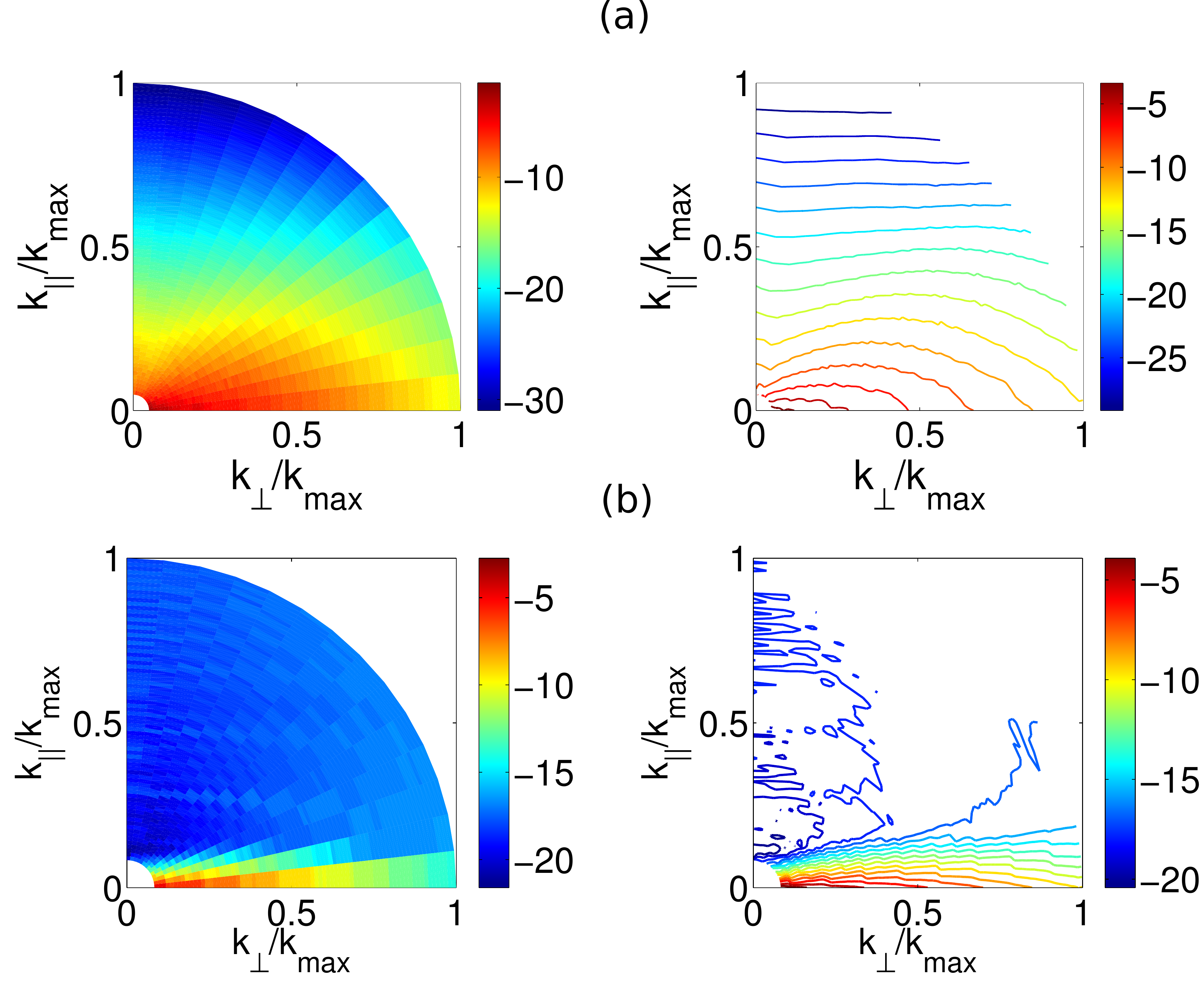}
\end{center}
\caption{ Density (left) and contour (right) plots of the energy spectrum for: (a) $N=18$ and (b) $N=130$.}
\label{fig:ring_spectrum}
\end{figure}

 The  external magnetic field induces a strong anisotropy in the flow.  A systematic study of anisotropic energy spectrum for various $N$'s have been presented in Reddy and Verma.\cite{Reddy:POF2014}  In Fig.~\ref{fig:ring_spectrum}, we exhibit the density and contour plots of the energy spectra for $N=18$ and 130.  These figures illustrate the energy concentrated near the equator,\cite{Caperan:JDM1985,Burattini:PD2008,Potherat:JFM2010} but there is a significant energy away from the equator.  This is the essential nature of quasi two-dimensional quasi-static MHD at high interaction parameters. 

In the next subsection, we will investigate how energy exchange takes place among the Fourier modes.
 
\subsection{Shell-to-shell energy transfers}\label{subsec:shell}

\begin{figure}[htbp]
\begin{center}
\includegraphics[width=8.3cm,angle=0]{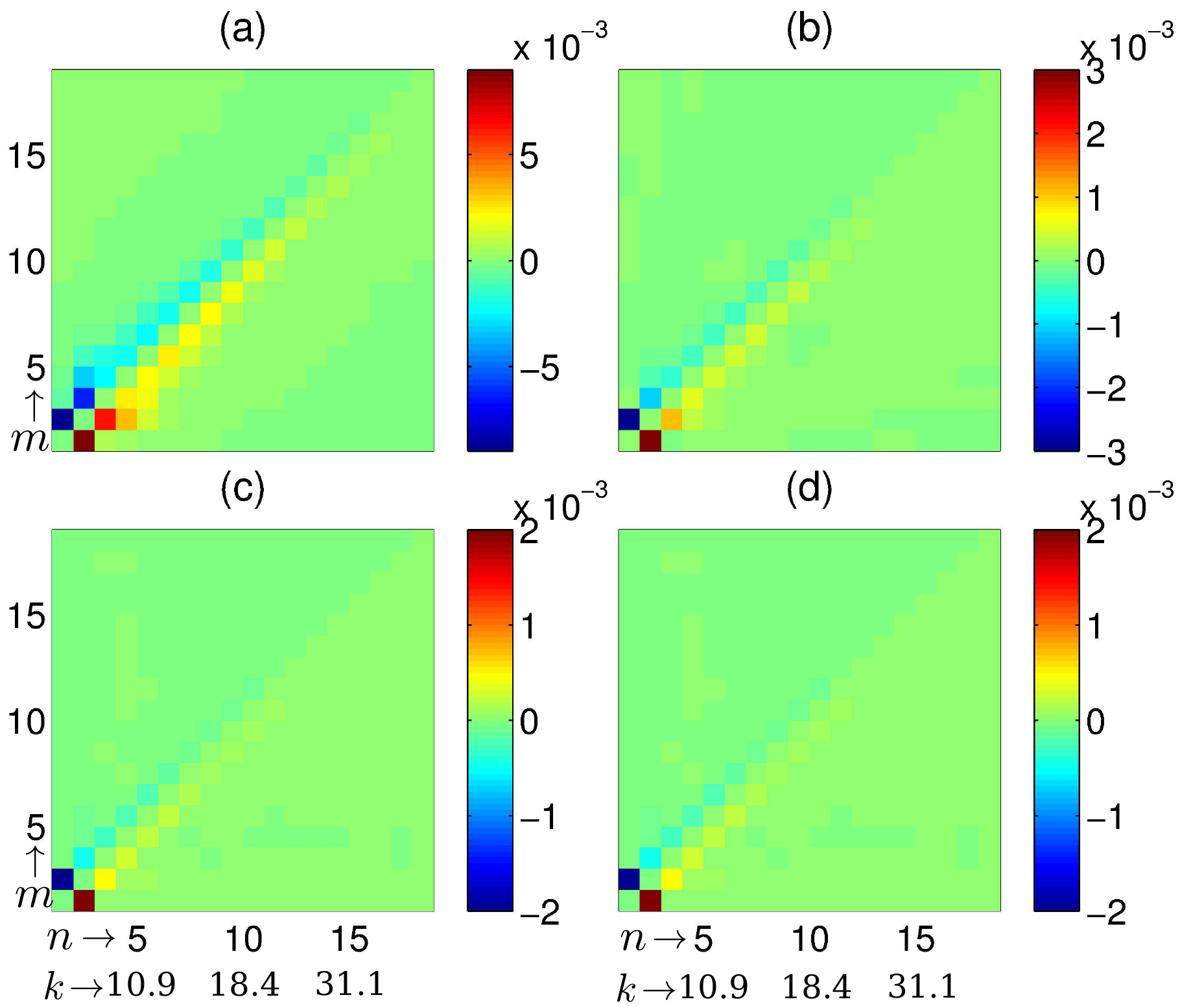}
\end{center}
\caption{Forward and local shell-to-shell energy transfer rates $T^m_n$ for: (a) $N=1.7$, (b) $N=11$, (c) $N=18$, and (d) $N=130$. Here, $m$ and $n$ are the giver and receiver shells, respectively, and $k$ is  the wavenumber of the outer radius of the corresponding shell.} 
\label{fig:shell}
\end{figure}

In Fig.~\ref{fig:shell}, we present the shell-to-shell energy transfer rates for $N=1.7, 11, 18$, and $130$. We observe that the $n^\mathrm{th}$ shell gives energy to the $(n+l)^\mathrm{th}$ shells ($l>0$), and it receives energy from the $(n-l)^\mathrm{th}$ shells. Thus, the shell-to-shell energy transfer for quasi-static MHD is forward.  We also observe that the maximum energy transfer is to the nearest neighbor, i.e., the $n^{\rm th}$ shell gives maximum positive energy transfer to the $(n+1)^\mathrm{th}$ shell, and maximum negative energy  to the $(n-1)^\mathrm{th}$ shell.  Hence, the shell-to-shell energy transfer is also local.  Our results are consistent with those  of Burattini {\it et al.}\cite{Burattini:PD2008}

\subsection{Ring-to-ring energy transfers}\label{subsec:ring}
\begin{figure}
\begin{center}
\includegraphics[width=8.5cm,angle=0]{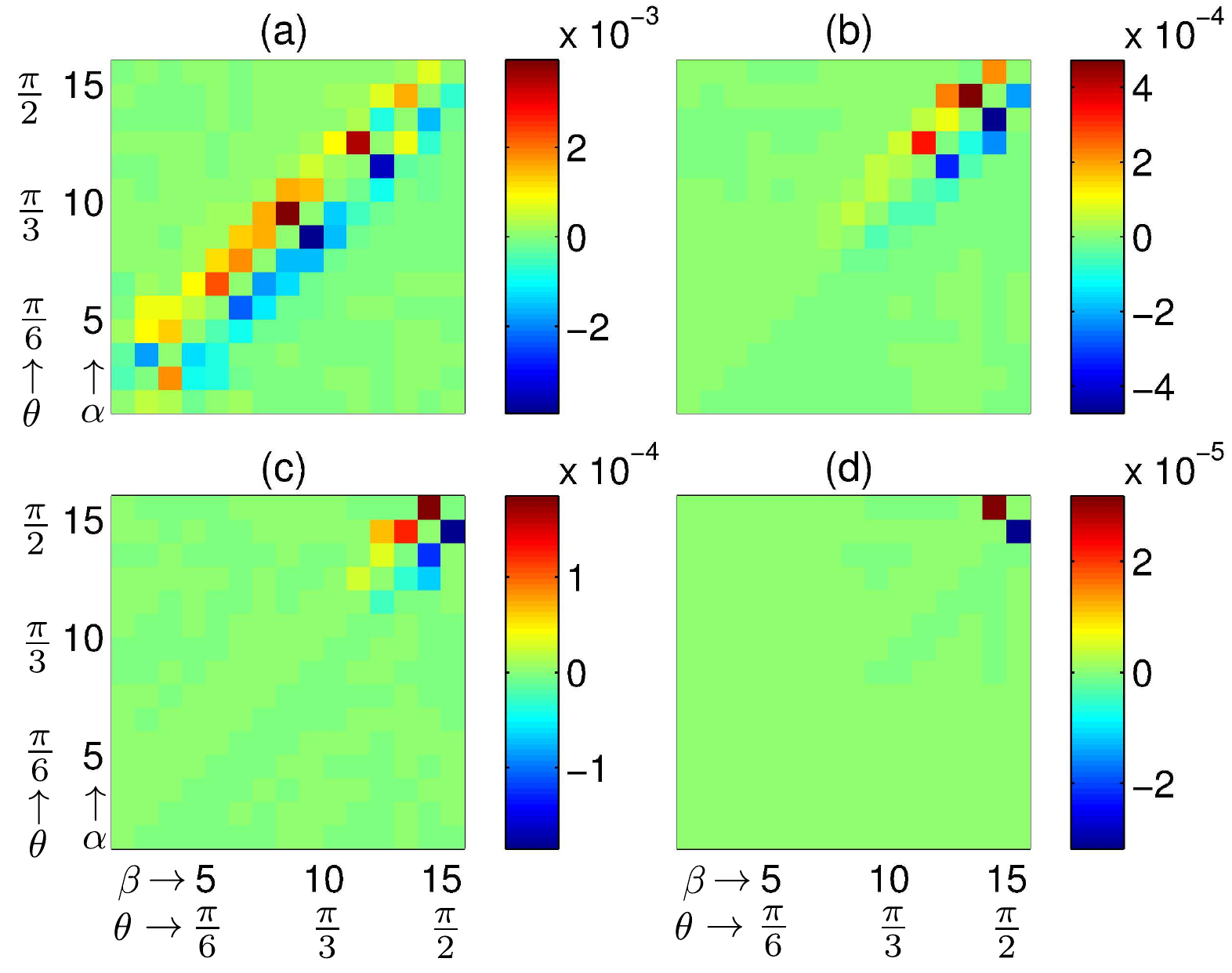}
\end{center}     
\caption{Ring-to-ring energy transfers $\overline{T}_{(9,\beta)}^{(9,\alpha)}$ among various rings of the $9^{\rm th}$ shell for: (a) $N=1.7$, (b) $N=11$, (c) $N=18$ and (d) $N=130$. Here,  $\alpha$ and $\beta$ are the indices of giver and receiver  rings,  respectively, and $\theta$  is the angle of the corresponding ring.  $\overline{T}_{(9,\beta)}^{(9,\alpha)}$ are dominant for neighboring rings  (local).  For large $N$, the energy transfers are dominant near the equator.}
\label{fig:ring-2-ring9_9}
\end{figure}

\begin{figure}
\begin{center}
\includegraphics[width=8.5cm,angle=0]{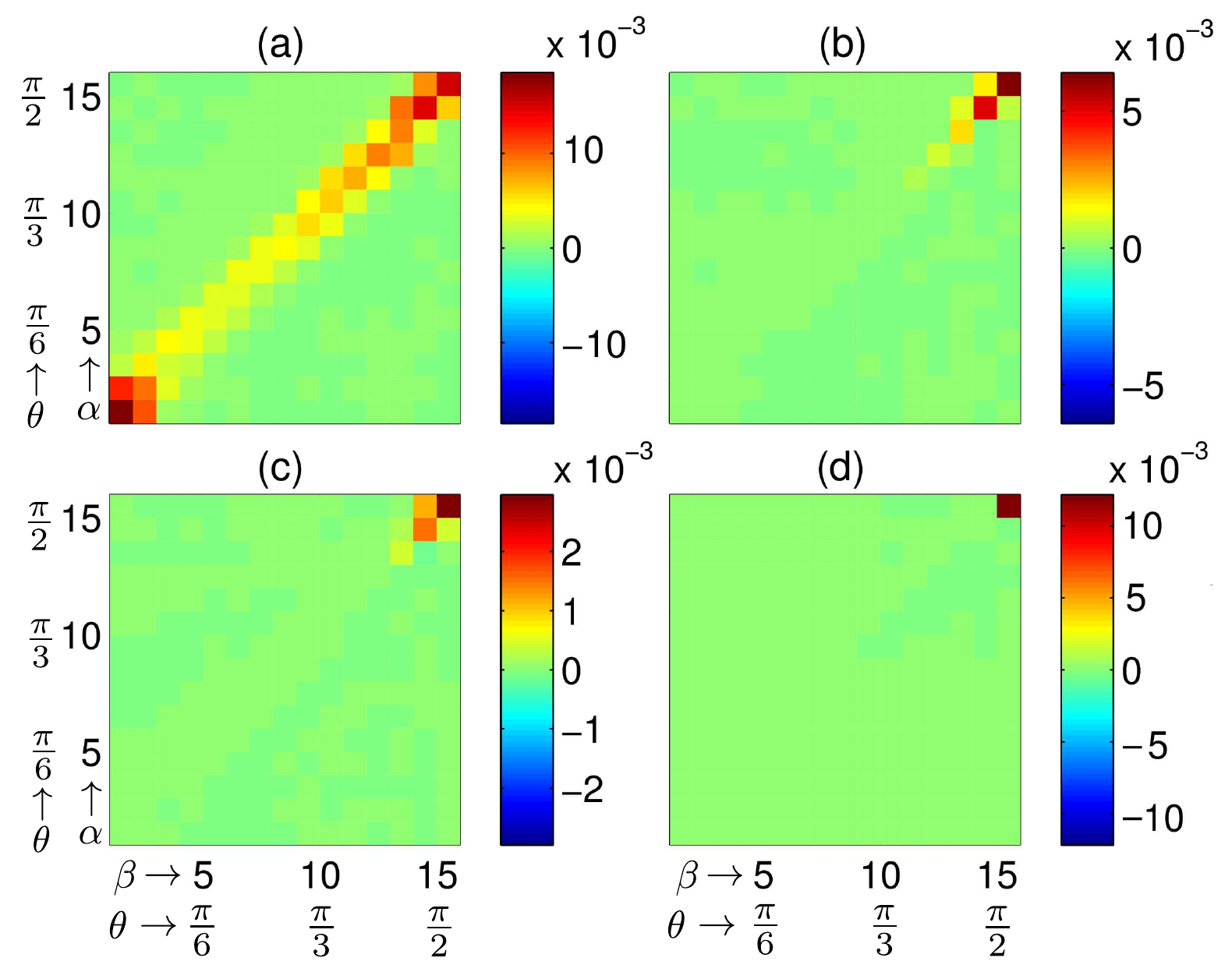}
\end{center}     
\caption{Local ring-to-ring energy transfers $\overline{T}_{(10,\beta)}^{(9,\alpha)}$ from the  rings of the $9^{\rm th}$ shell and to the rings of the $10^{\rm th}$ shell for: (a) $N=1.7$, (b) $N=11$, (c) $N=18$ and (d) $N=130$.   Note that $\overline{T}_{(10,\beta)}^{(9,\alpha)} > 0$.} 
\label{fig:ring-2-ring9_10}
\end{figure}

\begin{figure}
\begin{center}
\includegraphics[width=8.5cm,angle=0]{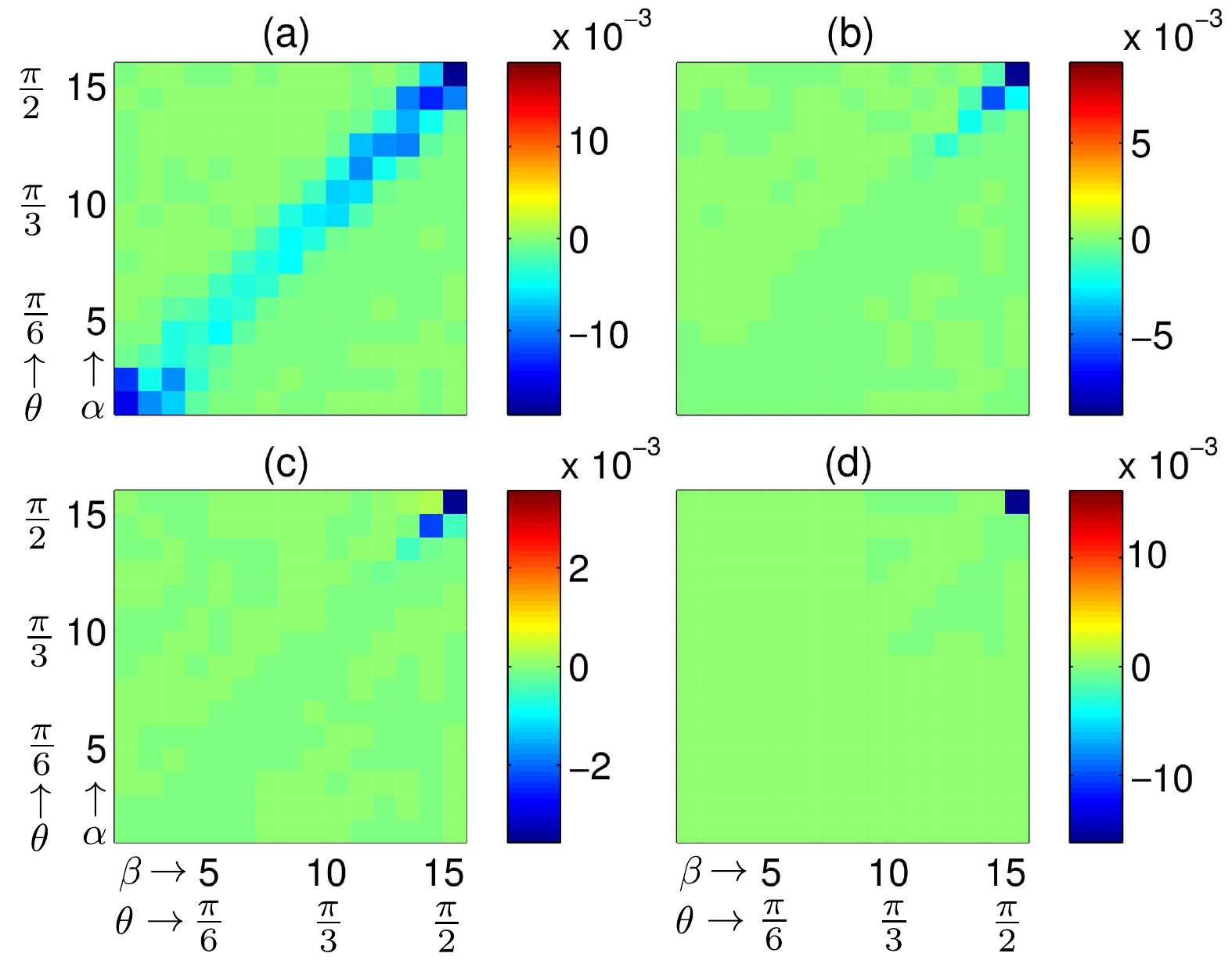}
\end{center}     
\caption{Local ring-to-ring energy transfers $\overline{T}_{(8,\beta)}^{(9,\alpha)}$ from the rings of the $9^{\rm th}$ shell and to the rings of the $8^{\rm th}$ shell for:  (a) $N=1.7$, (b) $N=11$, (c) $N=18$ and (d) $N=130$.   Note that $\overline{T}_{(8,\beta)}^{(9,\alpha)}<0$.}
\label{fig:ring-2-ring9_8}
\end{figure}

The angular dependence of the energy transfers can be computed using the ring-to-ring transfers. In Figs.~\ref{fig:ring-2-ring9_9}, \ref{fig:ring-2-ring9_10} and~\ref{fig:ring-2-ring9_8},  we illustrate the normalized ring-to-ring energy transfers $\overline{T}^{(m,\alpha)}_{(n,\beta)}$ from the rings of the $9^{\rm th}$ shell ($m=9$) to the rings of the shells $n=9,10$, and 8, respectively.   This analysis has been performed for $N=1.7, 11, 18$, and 130.   In these figures, the vertical axis represents the sector index of the giver ring ($\alpha$), while the horizontal axis represents the sector index for the receiver ring ($\beta$). 

First, we discuss $\overline{T}^{(9,\alpha)}_{(9,\beta)}$, i.e., the energy transfers among the rings with shell index 9.  Figure~\ref{fig:ring-2-ring9_9} shows that the energy transfer from  the ring $\alpha$ to the ring $(\alpha-1)$ is positive ($\overline{T}^{(9,\alpha)}_{(9,\alpha-1)} > 0$), while that from the ring $\alpha$ to the ring $(\alpha+1)$ is negative ($\overline{T}^{(9,\alpha)}_{(9,\alpha+1)} < 0$). Hence, the ring-to-ring energy transfer within a shell is from the equatorial region to the polar region.   Among the rings, the most significant energy transfers occur between the neighboring rings,  i.e., from a ring with index $\alpha$  to the rings with index  $\alpha\pm 1$. Hence, the energy transfer is local in the angular direction as well.   Another important conclusion that can drawn from the above computation is that for large $N$ ($N=11, 18, 130$), the dominant energy transfers takes place from the rings closer to the equator to  their neighbors (lower $\theta$).

Figure~\ref{fig:ring-2-ring9_10} illustrates $\overline{T}^{(9,\alpha)}_{(10,\beta)}$, i.e., the energy transfers from the rings in the $9^{\rm th}$ shell to those in the $10^{\rm th}$ shell.  The figure shows that  $\overline{T}^{(9,\alpha)}_{(10,\beta)} > 0$, and that they are most dominant for the equatorial rings ($\alpha, \beta \approx 15$).  Since $\overline{T}^{(9,\alpha)}_{(10,\beta)}$ dominates for $\alpha=\beta$, we conclude that the energy is transferred dominantly along a sector near the equator.  Hence, the energy transfers are forward along the sectors as well.  This feature is reinforced by $\overline{T}_{(8,\beta)}^{(9,\alpha)}$,  illustrated in  Fig.~\ref{fig:ring-2-ring9_8}, where we observe a negative energy being transferred
diagonally from the rings of shell 9 to the rings of shell 8.   Thus, the ring-to-ring transfers are local and forward.  For large $N$, these transfers tend to be dominant near the equator.

In the next subsection, we will describe conical energy flux.

\subsection{Conical Energy Flux} \label{subsec:conical}
\begin{figure}[htbp]
\begin{center}
\includegraphics[width=8.5cm,angle=0]{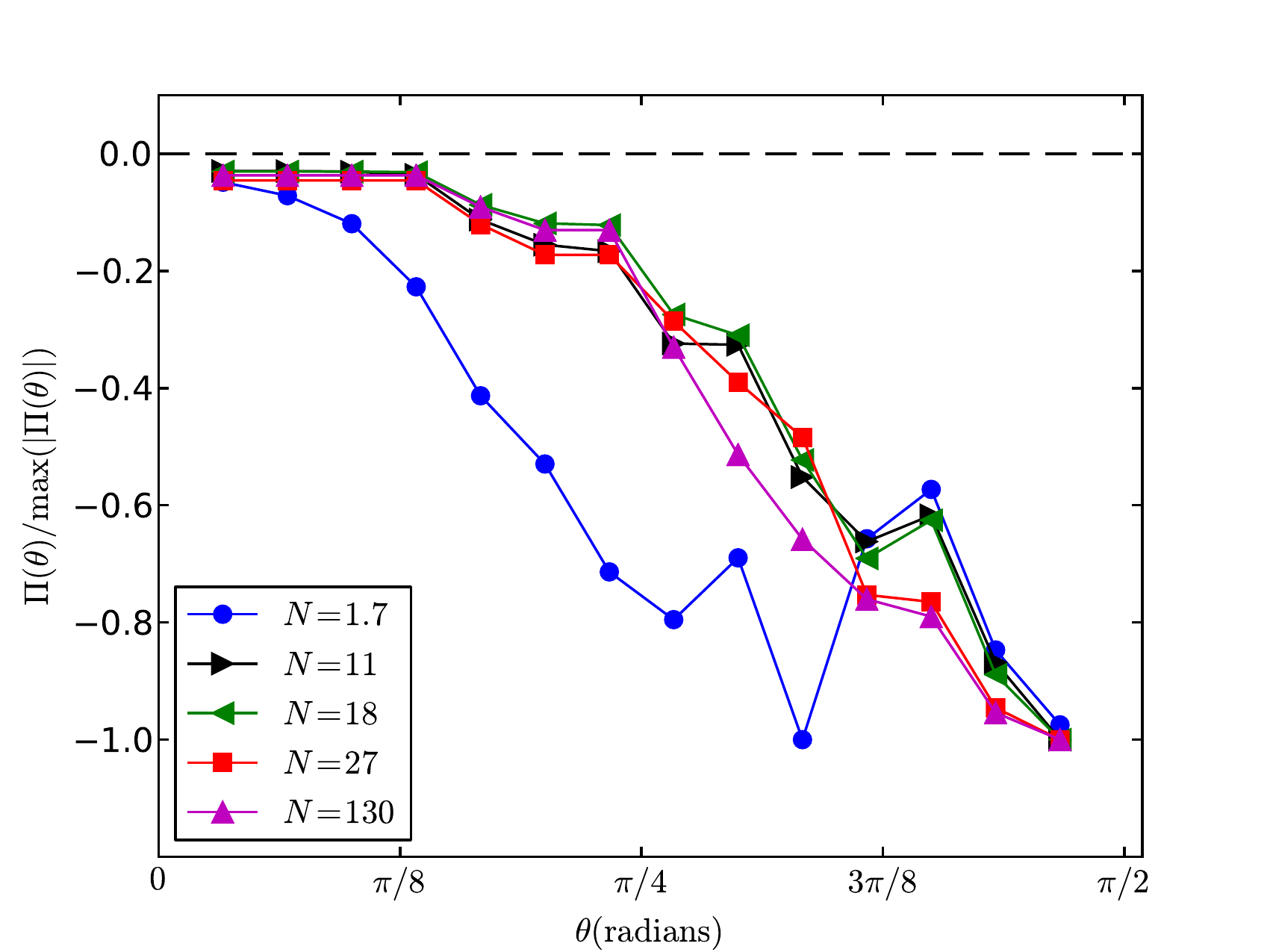}
\end{center}
\caption{Normalized conical energy flux, $\Pi({\theta})/\mathrm{max}(|\Pi({\theta})|)$, coming out of a cone of semi-vertical angle $\theta$ as a function of $\theta$ for various $N$'s.}
\label{fig:ang-flux}
\end{figure}

We can integrate the ring energy transfers over sectors and compute the conical energy flux [see Eq.~(\ref{eq:conical-flux})].  This  quantity describes the energy flux leaving a cone in the Fourier space (see Fig.~\ref{fig:flux-cone}). In Fig.~\ref{fig:ang-flux}, we plot the normalized flux $\Pi({\theta})/\mathrm{max}(|\Pi({\theta})|)$.  The figure shows that for $N=1.7$ to 130, the above flux is negative, indicating that the energy is transferred from the modes outside the cone to the modes inside the cone.  Note that $\Pi({\theta})/\mathrm{max}(|\Pi({\theta})|)$ is monotonic,  except for $N=1.7$ (due to the relatively weak magnetic field).   We also observe that the  maximal energy transfer takes place for the cone with a semi-vertical angle $\theta \approx \pi/2$.  Hence, the modes near the equatorial region transfer maximal energy towards the regions of smaller $\theta$. This energy gets dissipated by Joule heating, as well as it trickles down to the polar region.  

\begin{figure}[htbp]
\begin{center}
\includegraphics[width=8.5cm,angle=0]{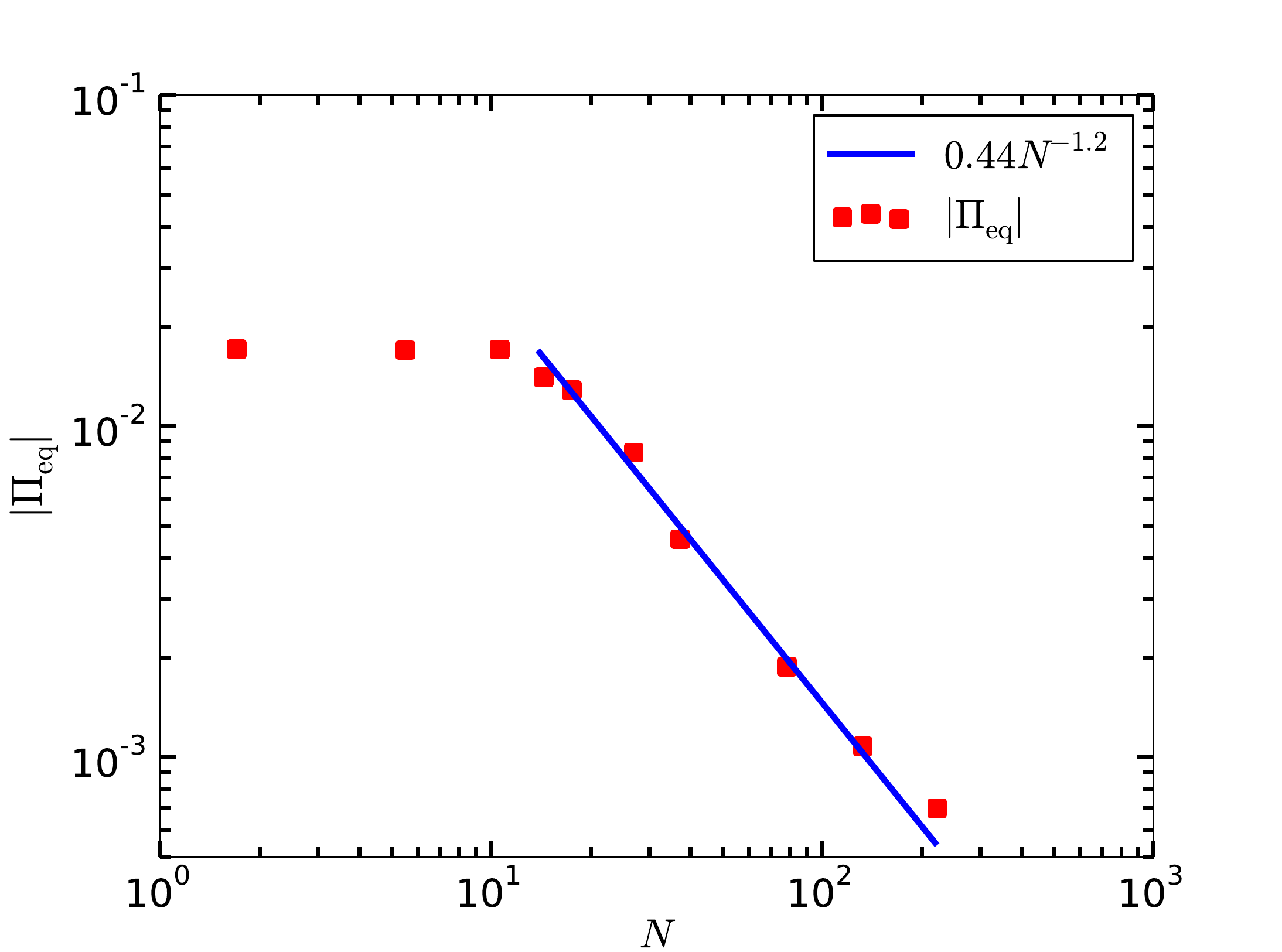}
\end{center}
\caption{Plot of $|\Pi_\mathrm{eq}|$ coming out of the equatorial sector as a function of $N$.  $|\Pi_\mathrm{eq}| \sim \mathrm{const.}$ for small $N$, but $|\Pi_\mathrm{eq}| \sim N^{-1.2}$ for $N>10$.  }
\label{fig:scale-ang-flux}
\end{figure}

In Fig.~\ref{fig:scale-ang-flux}, we plot the net energy transferred from the  cone with the largest  semi-vertical angle
\begin{equation}
\Pi_\mathrm{eq} = \sum_{\theta_p < \frac{7\pi}{15}} \sum_{\theta_k \ge \frac{7\pi}{15}}  {S({\bf k|p|q)}}. \label{eq:eqt-conical-flux}
\end{equation}
The quantity $-\Pi_{\rm eq}$ quantifies the energy transfer from the equatorial region to the modes inside the largest cone.  The figure indicates that $|\Pi_\mathrm{eq}|$ decreases very sharply with $N$ and follows  $|\Pi_\mathrm{eq}(N)| \propto N^{-1.2}$.

\begin{figure}[htbp]
\begin{center}
\includegraphics[width=8.5cm,angle=0]{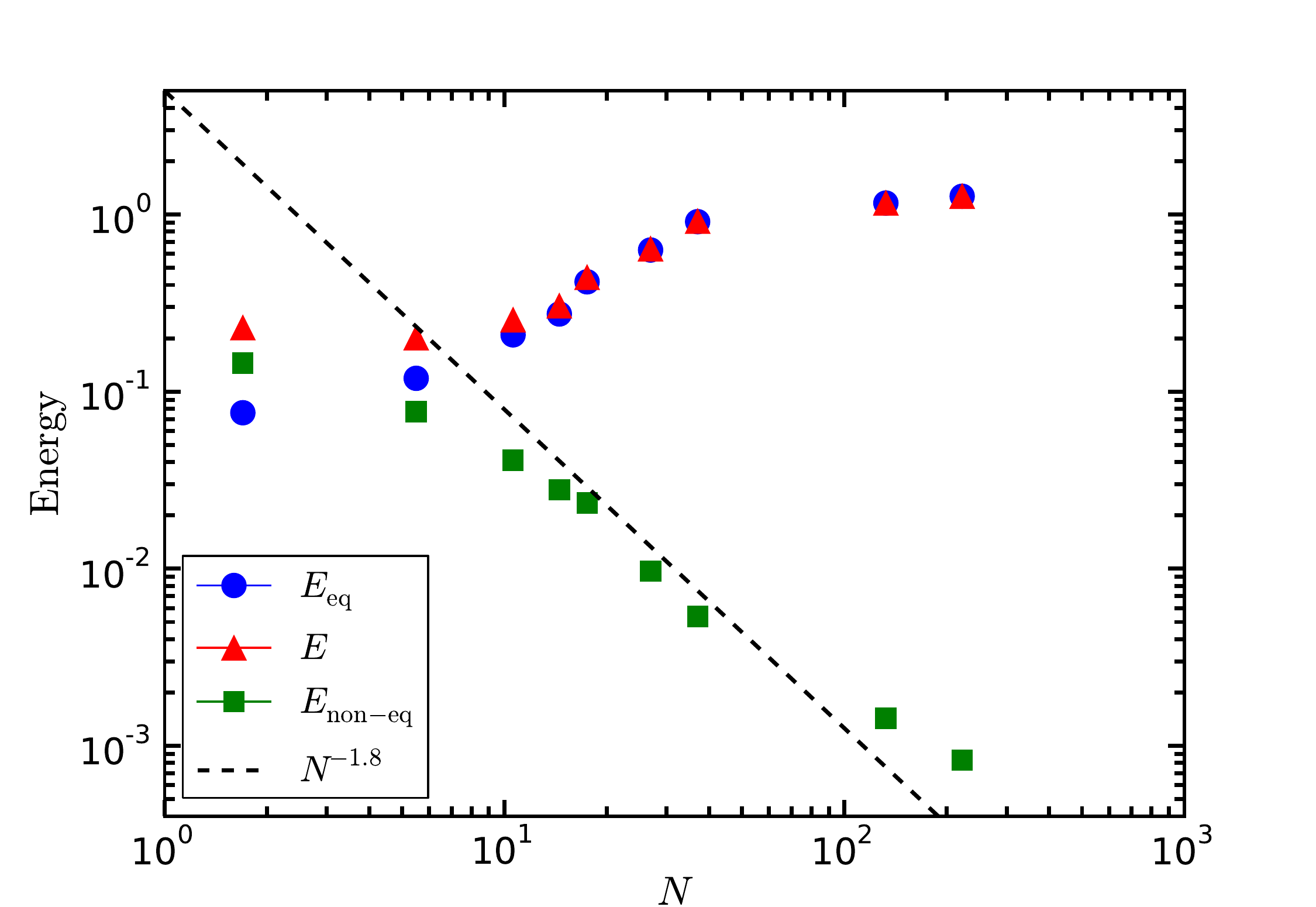}
\end{center}
\caption{Plots of $E, E_\mathrm{eq}, E_{\operatorname{non-eq}}$ vs. $N$.   $E_{\operatorname{non-eq}} \sim N^{-1.8}$ for $N>10$.}
\label{fig:E_eq}
\end{figure}

The decrease in $|\Pi_\mathrm{eq}|$ can be understood qualitatively using the energy distribution in the Fourier space.  In Fig.~\ref{fig:E_eq}, we plot the total energy and  energy contained in the equatorial region. The remaining energy, $ E_{\operatorname{ non-eq}} = E - E_{\rm eq}$, is also plotted in the figure.  We find that  $E_{\operatorname{non-eq}}$ decreases sharply with $N$ ($E_{\operatorname{non-eq}} \propto N^{-1.8}$).  Since the energy flux is a sum of $E(\mathbf p) E(\mathbf q)$, $E(\mathbf k) E(\mathbf p)$ and $E(\mathbf k) E(\mathbf q)$, apart from some other  factors (here $\mathbf k = \mathbf p + \mathbf q$)\cite{Lesieur:Book,Verma:PR2004}, and the receiver energy spectrum $E_{\operatorname{non-eq}} \propto N^{-1.8}$, it is reasonable that the conical energy flux $|\Pi_\mathrm{eq}|$ decreases very sharply.  Thus, we provide a qualitative explanation for the sharp decline of $|\Pi_\mathrm{eq}|$ with the interaction parameter.   This observation also explains why quasi-static MHD is quasi-two-dimensional for large $N$.  

\subsection{Energy fluxes of the parallel and perpendicular components} \label{subsec:perp_parallel}

Many experiments\cite{Alemany:JMec1979,Kolesnikov:FD1974} and numerical simulations\cite{Zikanov:JFM1998,Favier:POF2010b,Reddy:POF2014} indicate that quasi-static MHD exhibits quasi two-dimensional behavior for large $N$.  To probe the physics of energy transfers for large $N$ in detail,  we perform a numerical simulation for $N=100$ with  forcing applied at intermediate length scales ($8.0\leq  |{\bf k}_f| \leq 9.0$) to resolve the inverse and forward cascade regimes.   We take the final state of hydrodynamic simulation as an initial condition (see Sec.~\ref{sec:sim}) and apply an external magnetic field.  The simulation is carried out  till a final (quasi-steady) state is reached, which occurs at $t_{\rm final} \approx 400$.   The Joule dissipation, which is active at all scales, balances the energy growth due to the inverse cascade.

\begin{figure}[htbp]
\begin{center}
\includegraphics[width=8.5cm,angle=0]{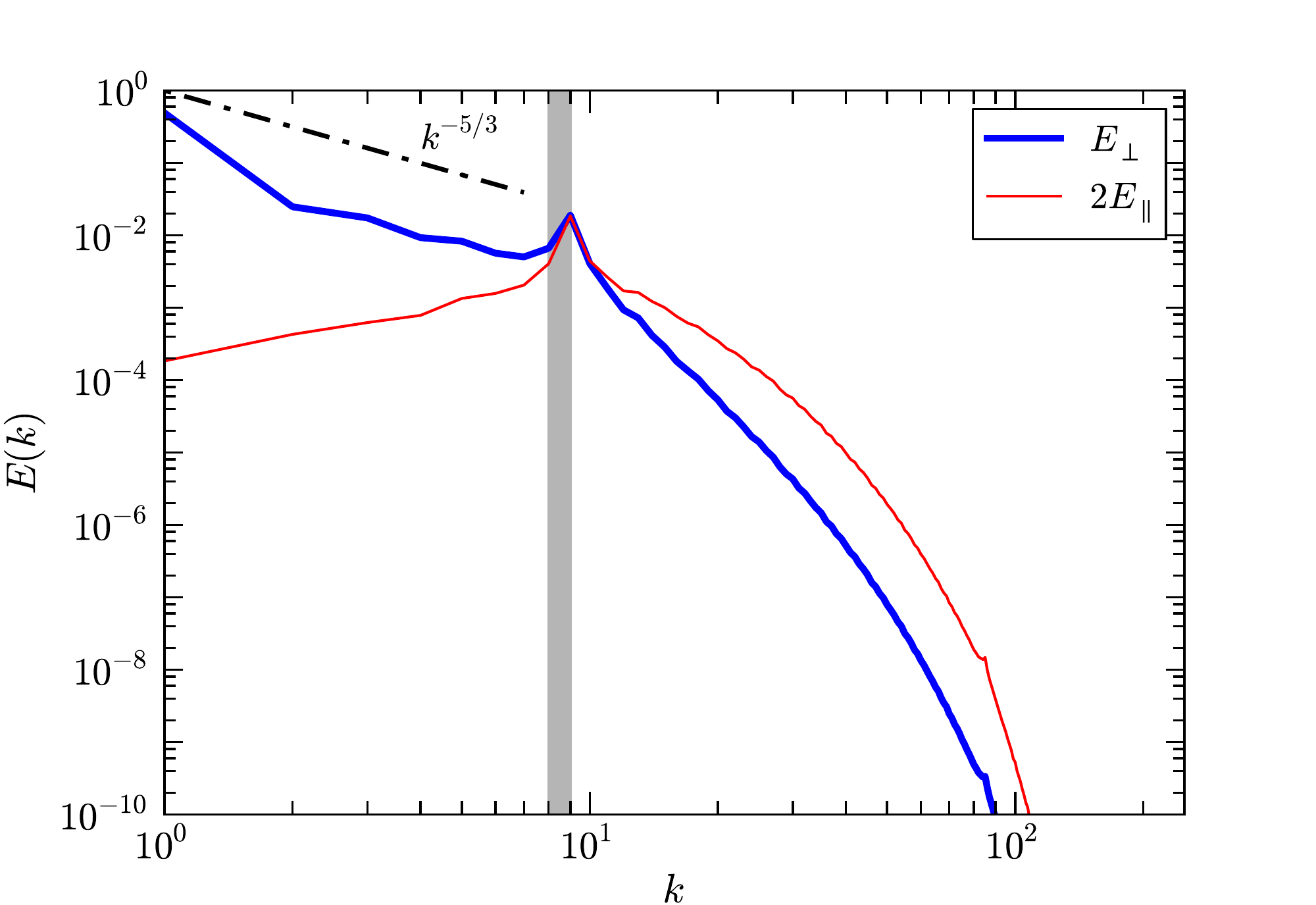}
\end{center}
\caption{ Plots of $E_{\perp}(k)$ and $2 E_{\parallel}(k)$ for $N=100$.   $E_{\perp}(k) > E_{\parallel}(k)$ for $k< k_f$, with $E_{\perp}(k)  \sim k^{-5/3}$, but $E_{\perp}(k) < E_{\parallel}(k)$ for $k> k_f$. The shaded region exhibits the forcing band $k_f \in [8,9]$.}
\label{fig:spec}
\end{figure}

In Fig.~\ref{fig:spec}, we plot the energy spectrum of the parallel and perpendicular components of the velocity field for $N=100$.  The figure  indicates that $E_\perp \gg 2E_\parallel $ for $k< k_f$, but $E_\perp \ll 2E_\parallel $ for  $k> k_f$.  We also observe that  $E_\perp(k)$ follows  $k^{-5/3}$ for $k<k_f$.  This feature demonstrates the quasi-two-dimensionalization of quasi-static MHD turbulence at high interaction parameters in periodic domains.  Our results are consistent with those of Favier {\it et al.}\cite{Favier:POF2010b}  To probe the physics of the flow further, we compute the energy fluxes of the parallel and perpendicular components of the velocity field.

\begin{figure}[htbp]
\begin{center}
\includegraphics[width=9.0cm,angle=0]{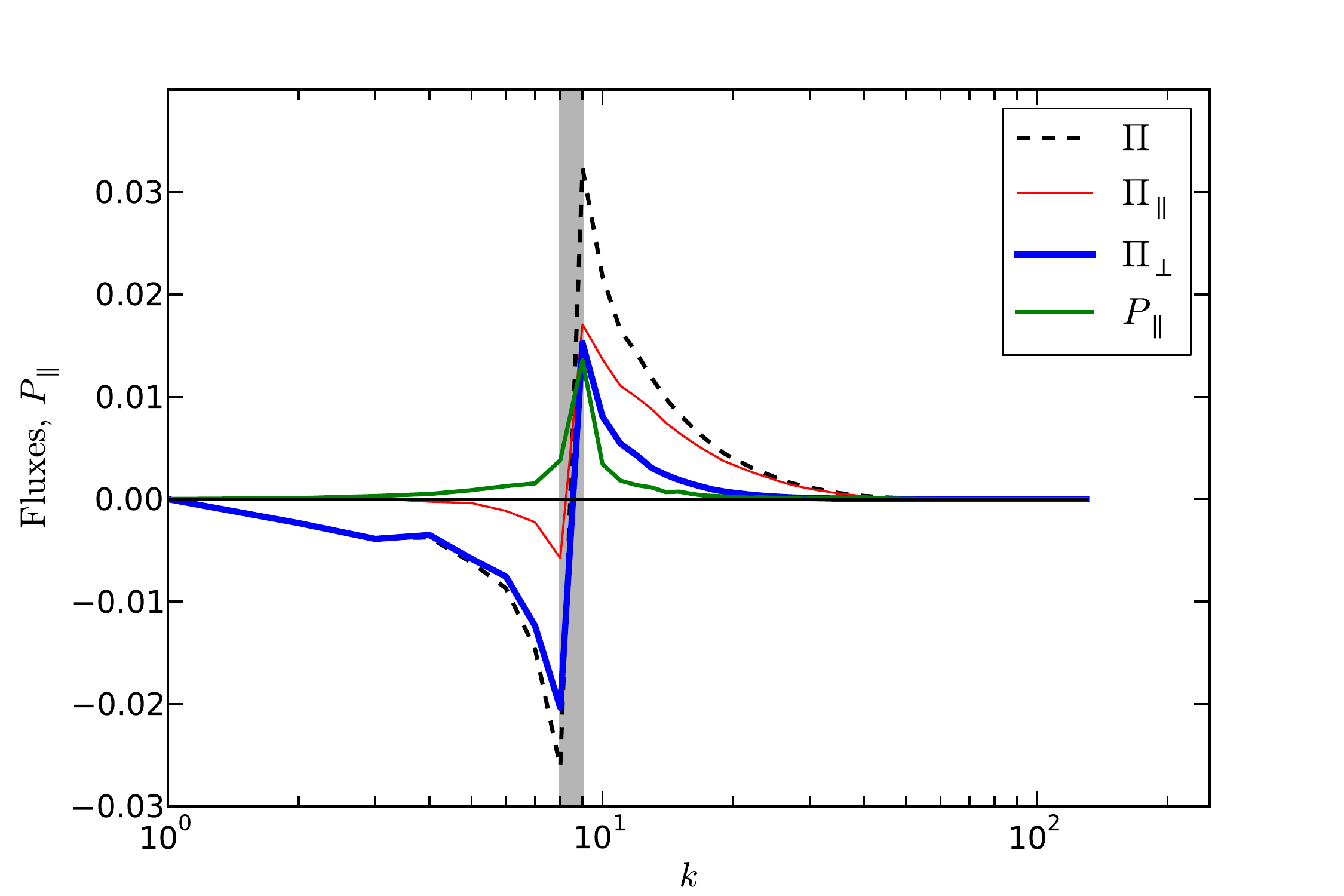}
\end{center}
\caption{Plots of the energy fluxes $\Pi(k)$,  $\Pi_{\parallel}(k)$,   $\Pi_{\perp}(k)$, and $P_\parallel(k)$ for $N=100$.   $\Pi_{\perp}(k) < 0$ for $k<k_f$, indicating an inverse cascade for ${\bf U}_\perp$, while $\Pi_\parallel(k) > 0$ for $k>k_f$, indicating a forward cascade for $U_\parallel$.  $P_\parallel(k) > 0$ for $k> k_f$, indicating an energy transfer from ${\bf U}_\perp$ to $U_\parallel$ via pressure. }
\label{fig:flux-par-perp}
\end{figure}

Figure~\ref{fig:flux-par-perp} exhibits the energy fluxes for the parallel and perpendicular components of the velocity field ($\Pi_\parallel$ and $\Pi_\perp$, respectively).     We observe that the $k<k_f$ and  $k>k_f$ regions are dominated by the $\Pi_\perp$ and $\Pi_\parallel$ fluxes, respectively.  The dominance of the negative energy flux for $\Pi_\perp$ in $k<k_f$ is consistent with the dominance of the inverse cascade of $\mathbf U_\perp$, while  $\Pi_\parallel>\Pi_\perp > 0$ in the $k>k_f$ region indicates the dominance of the forward cascade for $ U_\parallel$.  The aforementioned energy flux computations are consistent with the simulation results that $E_{\perp}(k) \gg E_{\parallel}(k) $ for lower wavenumbers, and $E_{\perp}(k) \ll E_{\parallel}(k) $  for higher wavenumbers (see Fig.~\ref{fig:spec}), which is consistent with the quasi-two-dimensional nature of quasi-static MHD turbulence at high interaction parameters.

In Fig.~\ref{fig:flux-par-perp}, we also plot $P_\parallel(k)$, which is the energy transferred to $ U_\parallel(k)$ from $\mathbf U_\perp(k)$ via pressure.   We observe that $P_\parallel(k)$ is positive for $k \ge k_f$.  Hence, $ U_\parallel(k)$ receives energy from $\mathbf U_\perp(k)$, which is consistent with the nature of the energy fluxes $\Pi_\parallel$ and $\Pi_\perp$ described above.  
 
\subsection{Dissipation rates}

The aforementioned preferential energy transfer from the equatorial region to the polar region can be understood using the distribution of the Joule dissipation $\epsilon_J$, which is proportional to  $(\cos^2\theta ) E({\bf k})$ [see Eq.~(\ref{eq:JD})].  Clearly, $\epsilon_J$ vanishes at the equatorial plane, where $\theta=\pi/2$.  However, $E({\bf k})$ increases monotonically with $\theta$.\cite{Potherat:JFM2010,Reddy:POF2014}   As a result, the Joule dissipation $\epsilon_J$ reaches  a maximum near  $\theta \approx \pi/2$,  but not at $\theta=\pi/2$ itself. To maintain a steady state, $\epsilon_J$ is balanced by a nonlinear energy transfer from the equatorial region.  This is the reason why the energy flows maximally from the equator towards the polar region (see Figs.~\ref{fig:ring-2-ring9_9} and~\ref{fig:ang-flux}). 

For large $N$,  $E({\bf k})$ is concentrated near the equator.  Therefore, $\epsilon_J$ peaks near $\theta=\pi/2$.  As a result, the ring-to-ring energy transfers are localized near the equator, as exhibited in Figs.~\ref{fig:ring-2-ring9_9}(c,d), \ref{fig:ring-2-ring9_10}(c,d), and \ref{fig:ring-2-ring9_8}(c,d).   These results are consistent with the quasi-two-dimensional behavior  of the quasi-static MHD flow for large $N$.\cite{Alemany:JMec1979, Zikanov:JFM1998,Favier:POF2010b}  

\begin{figure}[htbp]
\begin{center}
\includegraphics[width=8.5cm,angle=0]{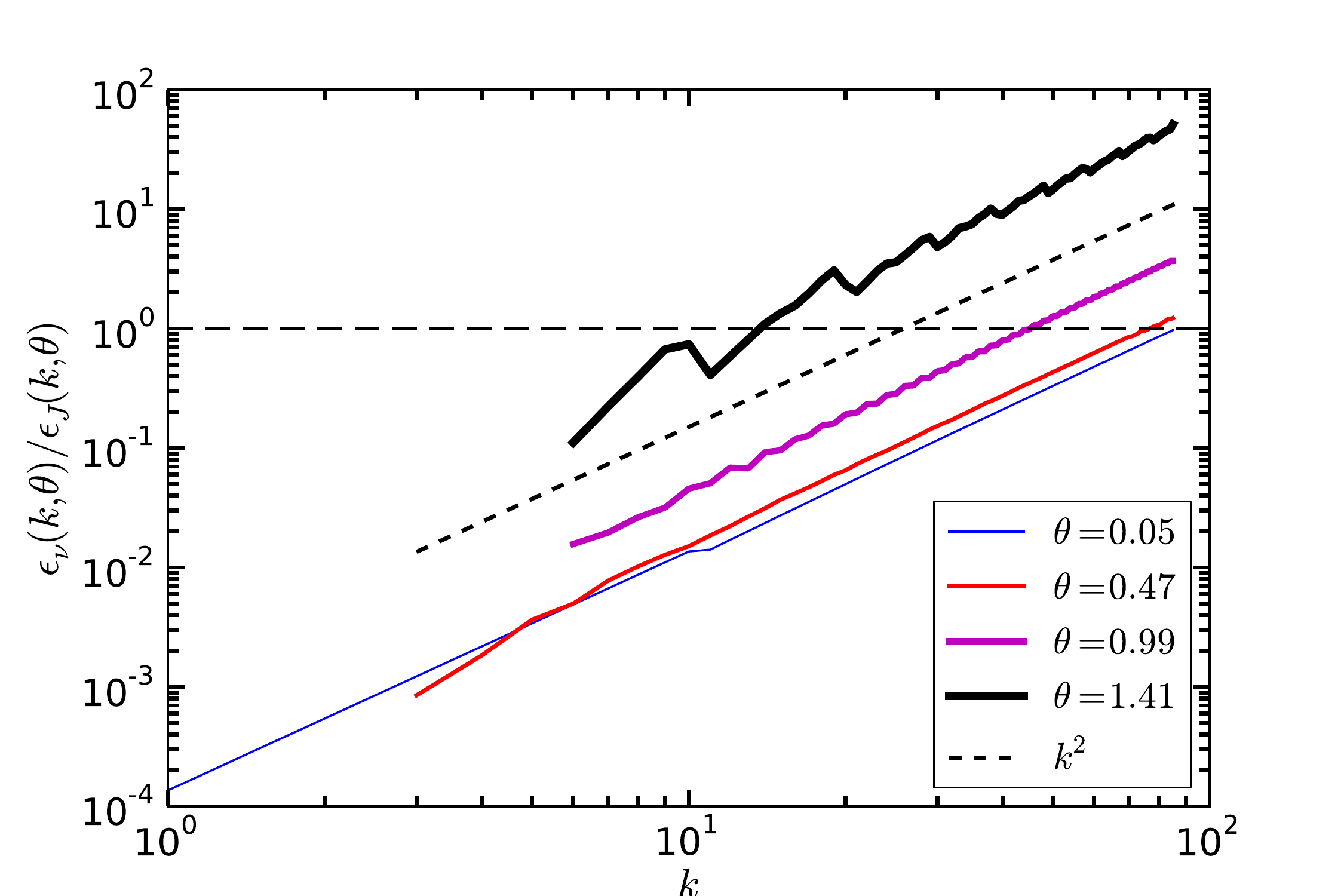}
\end{center}
\caption{For $N=27$,  $\epsilon_\nu(k,\theta)/\epsilon_J(k,\theta)$ vs.~$k$ for various sectors. $\epsilon_\nu(k,\theta)/\epsilon_J(k,\theta) \sim k^2$.  }
\label{fig:diss_spec_theta}
\end{figure}

Lastly we study the viscous and Joule dissipation rates for a large interaction parameter, here $N=27$.     Since
\begin{equation}
\frac{\epsilon_\nu(k,\theta)}{\epsilon_J(k,\theta)} = \frac{2 \nu' k^2 E(\mathbf k)}{2 B_0'^2 \cos^2\theta E(\mathbf k)} =  \frac{2 \nu' k^2 }{2 B_0'^2 \cos^2\theta },
\end{equation}
 $\epsilon_J(k,\theta)$ dominates $\epsilon_\nu(k,\theta)$ for 
\begin{equation}
 k < k_* = \frac{B_0' \cos\theta}{\sqrt{\nu'}},
 \label{eq:kstar}
\end{equation}
and vice versa.   This is expected, since the Joule dissipation is active at all wavenumbers, but the viscous dissipation acts strongly only at large wavenumbers.   In Fig.~\ref{fig:diss_spec_theta}, we plot $\epsilon_\nu(k,\theta)/\epsilon_J(k,\theta)$  as a function of the wavenumber $k$ for various sectors.  The mean angles of the chosen sectors are $\theta=0.05, 0.48, 0.99$, and 1.41.  

For a given sectorial angle $\theta$, the ratio $\epsilon_\nu(k,\theta)/\epsilon_J(k,\theta) \propto k^2$ because the viscous dissipation is proportional to $k^2$.  Consequently,  the Joule dissipation dominates at small wavenumbers, but the viscous dissipation takes over at large wavenumbers.  For a given wavenumber $k$, the ratio $\epsilon_\nu(k,\theta)/\epsilon_J(k,\theta) \propto 1/\cos^2\theta$; or, $\epsilon_\nu(k,\theta) \gg \epsilon_J(k,\theta)$ for the equatorial region ($\theta \approx \pi/2$) and vice versa for the polar region ($\theta \approx 0$).  The figure also indicates that the transition wavenumber $k_*$ decreases with increasing $\theta$, which is consistent with Eq.~(\ref{eq:kstar}).

\begin{figure}[htbp]
\begin{center}
\includegraphics[width=6.5cm,angle=0]{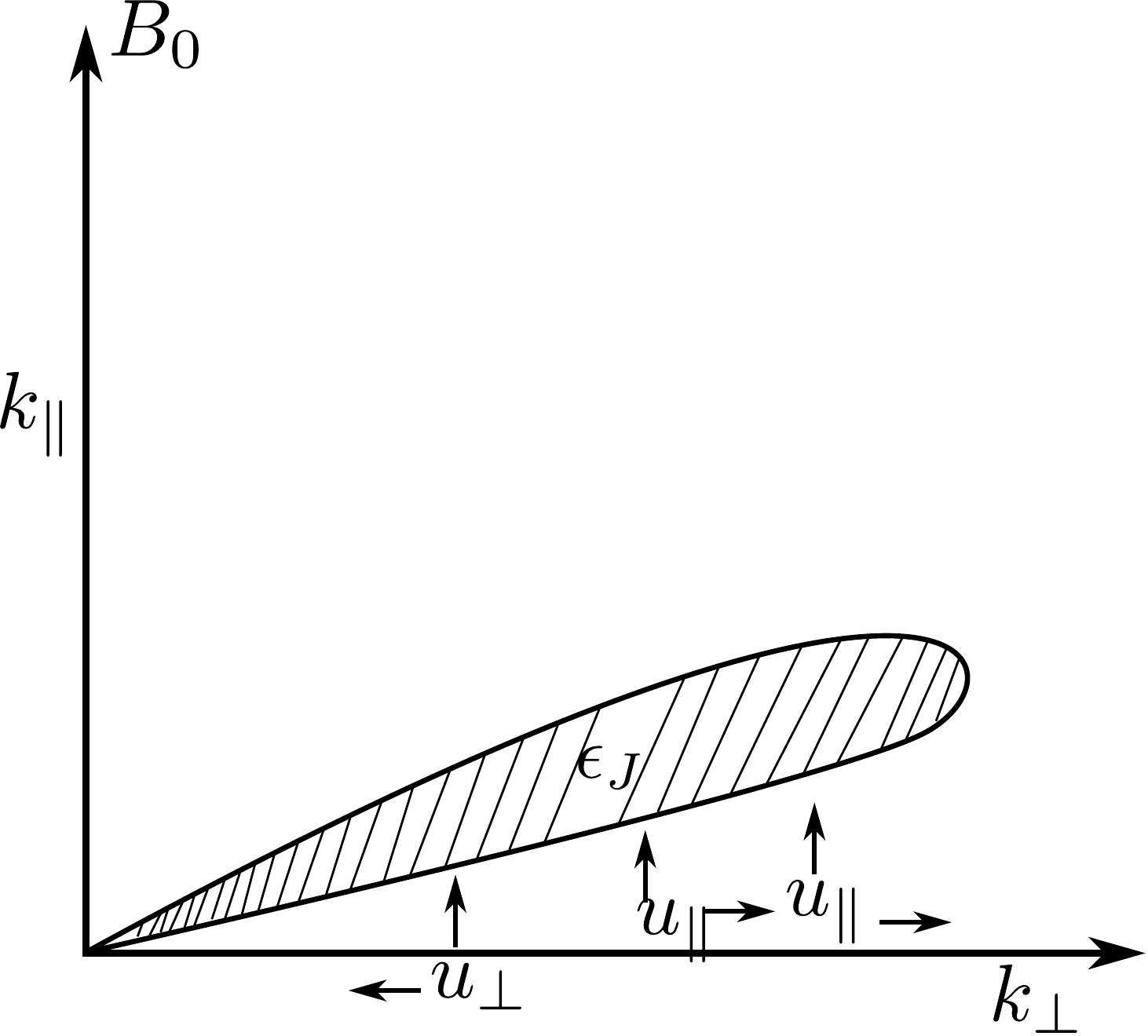}
\end{center}
\caption{A schematic illustration of the energy transfers (depicted by arrows) and dissipation rates in quasi-static MHD turbulence for large $N$.  ${\bf U}_\perp$ exhibits an inverse cascade, while $U_\parallel$ a forward cascade. }
\label{fig:mech}
\end{figure}

Our results are schematically illustrated in Fig.~\ref{fig:mech}.  The energy of the perpendicular component of the velocity cascades to smaller wavenumbers, while the energy of  the parallel component cascades to larger wavenumbers, where it gets depleted by the Joule dissipation via energy cascades to the polar region.

\section{Conclusions}
\label{sec:summary}
Earlier experiments and numerical simulations revealed that quasi-static MHD exhibits quasi-two-dimensional behavior  at high interaction parameters.\cite{Favier:POF2010b,Zikanov:JFM1998,Alemany:JMec1979}   In this paper, we have studied the energy transfer mechanisms operating in   quasi-static MHD and show them to be consistent with the aforementioned anisotropic energy distribution.  Here, we have studied the shell-to-shell and ring-to-ring energy transfers, as well as the conical flux.  We have also studied the energy fluxes of the parallel and perpendicular components of the velocity field.  For most of our runs, our forcing wavenumber band lies in the small-wavenumber regime.   

The main results of our paper are:

\begin{enumerate}
\item We have developed a formalism to compute the conical energy transfer.  We also provided a scheme to compute the energy fluxes for the parallel and perpendicular components of the velocity field.

\item Earlier, Burattini {\it et al.} \cite{Burattini:PD2008} showed that the shell-to-shell energy transfer is local.  In this paper, we show that the ring-to-ring energy transfers are forward and  local, both in wavenumber shells and angles. Within a shell, the ring-to-ring transfers are from higher polar angles to lower polar angles (i.e., from the equatorial region to the polar region).  For the rings across  shells, it is dominantly along the same sector or neighboring sectors. 

\item When the flow is forced at an intermediate wavenumber band, for large $N$,  we observe that the  inverse cascade at low wavenumbers is dominated by the negative energy flux of the perpendicular component of velocity, while the forward cascade at large wavenumbers is dominated by the positive energy flux of the parallel component.   
 
\end{enumerate}
 
In conclusion, the energy transfers in quasi-static MHD provide valuable insights into the physics of the flow.  The energy transfers in quasi-static MHD have similarities with full MHD, rotating, and stratified turbulence.  Hence the tools developed in the present paper may be useful for such studies.

\acknowledgments
We thank  D.~Carati, B.~Knaepen, B.~Teaca, P.~Perlekar, P.~Satyamurthy,  and D.~Biswas for useful discussions and the anonymous referee for helpful comments. RK thanks S.~V.~G.~Menon, former Head, Theoretical Physics Division BARC for the encouragement and support.  This work was supported by Board of Research in Nuclear Science, Department of Atomic Energy, Govt. of India through research grant 2009/36/81-BRNS. All the simulations were performed on the {\it {HPC system}} and {\it Chaos} cluster of IIT Kanpur.

\section*{Appendix A: Mode-to-mode energy transfers for the perpendicular and parallel components of  the velocity field}\label{appendix}

In this appendix, we derive formulas for the energy transfers for the perpendicular and parallel components of the velocity field.  We focus on a triad $({\bf k,p,q})$ under the limit $\nu=0$ and $B_0=0$.  Note that ${\bf k'+p+q}= 0$, and ${\bf k'} = -{\bf k}$.

Following Dar {\it et al.}\cite{Dar:PD2001} and Verma,\cite{Verma:PR2004} we derive the following equations  from Eqs.~(\ref{eq:NS2},\ref{eq:continuity2}):
\begin{eqnarray}
{{\partial E_{\perp}({\bf k'})}\over {\partial t}} &=&   S_{\perp}({\bf k'}|{\bf p}|{\bf q}) + S_{\perp}({\bf k'}|{\bf q}|{\bf p}) +  P_{\perp}({\bf k'}) , \label{eq:dt_Eperp} \\
{{\partial E_{\parallel}({\bf k'})}\over {\partial t}} &=&   S_{\parallel}({\bf k'}|{\bf p}|{\bf q}) + S_{\parallel}({\bf k'}|{\bf q}|{\bf p}) +  P_{\parallel}({\bf k'}), \label{eq:dt_Epll} \
\end{eqnarray}
\noindent 
where $E_{\perp}({\bf k}) =E_{\perp}({\bf k'})  =\frac{1}{2}|\hat {\bf U}_{\perp}({\bf k})|^2$ and $E_{\parallel}({\bf k}) = E_{\parallel}({\bf k'})  = \frac{1}{2}|\hat { U}_{\parallel}({\bf k})|^2$ are the energies of the perpendicular and parallel components of the velocity field, respectively, and
\begin{eqnarray}
S_{\perp}({\bf k'|p|q})&=& -\mathrm{\Im} \{ [ {\bf k' \cdot {\hat U}(q)] [\hat U_{\perp}(k') \cdot \hat U_{\perp}(p)}]\},\\
S_{\parallel}({\bf k'|p|q})&=& -\mathrm{\Im} \{ [ {\bf k' \cdot {\hat U}({\bf q})}] [\hat U_{\parallel}({\bf k'})  \hat U_{\parallel}({\bf p})] \},\\
P_{\perp}({\bf k'}) &=& -\Im \{ [{\bf k' \cdot \hat U_{\perp}(k')} ] \hat{P}({\bf k'})\}, \\
P_{\parallel}({\bf k'}) &=& -\Im \{ [  k'_{\parallel}  \hat U_{\parallel}({\bf k'}) ] \hat{P}({\bf k'})\}, 
\end{eqnarray}
where $\Re,\Im$, * represent the real part and imaginary part, and the complex conjugate of a complex number, respectively.    Equations~(\ref{eq:dt_Eperp},\ref{eq:dt_Epll}) indicate that the mode ${\bf k^\prime}$ receives energy from modes {\bf p} and {\bf q}. Similarly we can also derive that 
\begin{eqnarray}
{{\partial E_{\perp}({\bf p})}\over {\partial t}} &=&   S_{\perp}({\bf p}|{\bf q}|{\bf k'}) + S_{\perp}({\bf p}|{\bf k'}|{\bf q}) +  P_{\perp}({\bf p}) ,\\
{{\partial E_{\parallel}({\bf p})}\over {\partial t}} &=&   S_{\parallel}({\bf p}|{\bf q}|{\bf k'}) + S_{\parallel}({\bf p}|{\bf k'}|{\bf q}) +  P_{\parallel}({\bf p}) ,\\
{{\partial E_{\perp}({\bf q})}\over {\partial t}} &=&   S_{\perp}({\bf q}|{\bf k'}|{\bf q}) + S_{\perp}({\bf q}|{\bf p}|{\bf k'}) +  P_{\perp}({\bf q}) ,\\
{{\partial E_{\parallel}({\bf q})}\over {\partial t}} &=&   S_{\parallel}({\bf q}|{\bf k'}|{\bf q}) + S_{\parallel}({\bf q}|{\bf p}|{\bf k'}) +  P_{\parallel}({\bf q}).
\end{eqnarray}
Using ${\bf k \cdot {\hat U}(k)}=0$, we can show that 
\begin{eqnarray}
P_{\perp}({\bf k'})+P_{\parallel}({\bf k'}) &=& 0,\\
S_{\perp}({\bf k'}|{\bf p}|{\bf q}) &=& -S_{\perp}({\bf p}|{\bf k'}|{\bf q}),\\
S_{\parallel}({\bf k'}|{\bf p}|{\bf q}) &=& -S_{\parallel}({\bf p}|{\bf k'}|{\bf q}).
\end{eqnarray}
Using the above, we can conclude that 
\begin{eqnarray}
{{\partial }\over {\partial t}} & & \left[ E_{\perp}({\bf k'}) + E_{\perp}({\bf p}) + E_{\perp}({\bf q}) \right]  \nonumber \\
&& =  P_{\perp}({\bf k'}) + P_{\perp}({\bf p})+P_{\perp}({\bf q}), \label{eq:Eperp} \\
{{\partial }\over {\partial t}} & & \left[ E_{\parallel}({\bf k'}) + E_{\parallel}({\bf p}) + E_{\parallel}({\bf q}) \right]  \nonumber \\
&& =  -\left[ P_{\perp}({\bf k'}) + P_{\perp}({\bf p})+P_{\perp}({\bf q}) \right]. \label{eq:Epll}
\end{eqnarray}

Therefore, we can make the following conclusions regarding the energy transfers for the parallel and perpendicular components of the velocity field:
\begin{enumerate}
\item The sum of Eqs.~(\ref{eq:Eperp}, \ref{eq:Epll}) shows that the total energy (sum of the perpendicular and parallel components) for a triad is conserved. However, there is an energy transfer between the perpendicular and parallel components via pressure.

\item The perpendicular component ${\bf \hat U_\perp(k')}$ receives  energy by an amount $S_{\perp}({\bf k'|p|q})$ from ${\bf \hat U_\perp(p)}$ with ${\bf \hat U(q)}$ as a mediator. Symmetrically, it also receives energy by an amount $S_{\perp}({\bf k'|q|p})$ from  ${\bf \hat U_\perp(q)}$ via ${\bf \hat U(p)}$.

The parallel component ${ \hat U_\parallel({\bf k')}}$ receives energy by amounts $S_{\parallel}({\bf k'|p|q})$ and $S_{\parallel}({\bf k'|q|p})$ from the modes ${ \hat U_\parallel({\bf p})}$ and ${ \hat U_\parallel({\bf q})}$, respectively (with ${\bf \hat U(q)}$ and ${\bf \hat U(p)}$ acting as the respective mediators).

\item  Equation~(\ref{eq:dt_Eperp}) implies that the perpendicular component ${\bf \hat U_\perp(k')}$ gains energy from the $P_{\perp}({\bf k'})$ term, which arises due to the pressure.  Since  $P_{\perp}({\bf k'}) = -P_{\parallel}({\bf k'}) $, the energy gained by ${\bf \hat U_\perp(k')}$ via pressure is the same as the energy lost by ${ \hat U_\parallel({\bf k})}$ (see Eq.~(\ref{eq:dt_Epll})).  Hence, the energy transfer between the parallel and perpendicular components occurs via pressure.
\end{enumerate}

We use these formulas to compute the energy fluxes of the perpendicular and parallel components of the velocity field.


%

\end{document}